\documentclass[letter]{aa}  
\usepackage{natbib}
\bibpunct{(}{)}{;}{a}{}{,} 
\usepackage{graphicx}
\usepackage{academicons}
\usepackage{color}
\usepackage{float}
\usepackage{flushend}
\usepackage{txfonts}
\usepackage{hyperref}
\usepackage{tablefootnote}
\def\fm{\hbox{$.\!\!^{\rm m}$}}

\newcommand{\gcc}{g\,cm$^{-3}$}

\graphicspath{{./}{figures/}}

\begin{document}
\titlerunning{Tidally locked rotation of Eris from ground based and space photometry}
\authorrunning{Szak\'ats et al.} 

\title{Tidally locked rotation of the dwarf planet (136199) Eris discovered from long-term ground based and space photometry}

\author{R. Szakáts \orcid{0000-0002-1698-605X} \inst{1,2} 
          \and
          Cs. Kiss \orcid{0000-0002-8722-6875} \inst{1,2,3}
          \and
          J. L. Ortiz \inst{6}
          \and
          N. Morales \inst{6}
          \and
          A. P\'al \orcid{0000-0001-5449-2467} \inst{1,2,4}
          \and
          T. G. M\"uller \orcid{0000-0002-0717-0462} \inst{5}
          \and
          J. Greiner \inst{5}
          \and
          P. Santos-Sanz \orcid{0000-0002-1123-983X} \inst{6}
          \and
          G. Marton \orcid{0000-0002-1326-1686} \inst{1}
          \and
          R. Duffard \inst{6}
          \and
          P. S\'agi \orcid{0000-0001-7549-0348} \inst{1,2,4}
          \and
          E. Forgács-Dajka\orcid{0000-0002-5735-6273}\inst{7,8,9,10}
          }
          
\institute{Konkoly Observatory, Research Centre for Astronomy and Earth Sciences, Konkoly Thege 15-17,               H-1121 Budapest, Hungary\\
           \email{szakats.robert@csfk.org}
         \and
           CSFK, MTA Centre of Excellence, Budapest, Konkoly Thege Miklós út 15-17, H-1121, Hungary
         \and
             ELTE Eötvös Loránd University, Institute of Physics, Budapest, Hungary
         \and
             E\"otv\"os Lor\'and University, P\'azm\'any P\'eter s\'et\'any 1/A, H-1171 Budapest, Hungary
         \and
             Max-Planck-Institut f\"ur extraterrestrische Physik, D-85748 Garching, Germany
         \and
             Instituto de Astrof\'\i{}sica de Andaluc\'\i{}a, IAA-CSIC, Glorieta de la Astronom\'\i{}a
             s/n, E-18008 Granada, Spain
        \and
            Department of Astronomy, Institute of Geography and Earth Sciences, E\"otv\"os Lor\'and University, H-1117 Budapest, P\'azm\'any P\'eter s\'et\'any 1/A, Hungary
        \and
            Centre for Astrophysics and Space Science, E\"otv\"os Lor\'and University, H-1117 Budapest, P\'azm\'any P\'eter s\'et\'any 1/A, Hungary
        \and    
            ELKH-SZTE Stellar Astrophysics Research Group, H-6500 Baja, Szegedi út, Kt. 766, Hungary
        \and
            Wigner Research Centre for Physics, P.O. Box 49, Budapest H-1525, Hungary
        }

\abstract{

The rotational states of the members in the dwarf planet - satellite systems in the transneptunian region are determined by the formation conditions and the tidal interaction between the components, and these rotational characteristics are the prime tracers of their evolution.  
Previously a number of authors claimed highly diverse values for the rotation period for the dwarf planet Eris, ranging from a few hours to a rotation (nearly) synchronous with the orbital period (15.8\,d) of its satellite, Dysnomia. In this letter we present new light curve data of Eris, taken with $\sim$1-2m-class ground based telescopes, and with the TESS and Gaia space telescopes. 
TESS data could not provide a well-defined light curve period, but could constrain light curve variations to a maximum possible light curve amplitude of $\Delta m$\,$\leq$\,0.03\,mag (1-$\sigma$) for P\,$\leq$\,24\,h periods. Both the combined ground-based data and the Gaia measurements unambiguously point to a light curve period equal to the orbital period of Dysnomia, P\,=\,15.8\,d, with a light curve amplitude of $\Delta m$\,$\approx$\,0.03\,mag, i.e. the rotation of Eris is tidally locked. Assuming that Dysnomia has a collisional origin, calculations with a simple tidal evolution model show that Dysnomia has to be relatively massive (mass ratio of q\,=\,0.01--0.03) and large (radius of $R_s$\,$\geq$\,300\,km) to slow down Eris to synchronized rotation. These simulations also indicate that 
-- assuming tidal parameters usually considered for transneptunian objects -- 
the density of Dysnomia should be 1.8-2.4\,\gcc, an exceptionally high value among similarly sized transneptunian objects, putting important constraints on the formation conditions.}


\maketitle

\section{Introduction} \label{Sect:intro}

The largest (D\,$\gtrsim$1000\,km) solar system objects -- the dwarf planets -- represent a separate class among transneptunian objects with distinct surface characteristics and internal properties and also with a high incidence of satellites \citep{Brown2006,Kiss2017}.
The present rotational state of these large bodies are expected to be a combined outcome of formation conditions and tidal interactions in the case of a massive satellite, and therefore their rotational light curves and the properties derived from them are important clues to unravel their history. 
Among these objects, on one side, the Pluto-Charon system is known to be tidally locked \citep{D97} while Haumea is an extremely fast rotator with a system of two satellites in which the more massive satellite could not reach rotational synchronization with its orbital period \citep{Hastings2016}. The rotation periods of other transneptunian dwarf planets range from a few hours to a few days \citep[e.g. Quaoar, Gonggong and Makemake;][]{Ortiz2003,Pal2016,Hromakina2019}, indicating a wide range of formation conditions and/or tidal interactions. For these objects with relatively long rotation periods the light curve is expected to be caused by albedo variegations on the surface instead of being spin-shape driven.  
Eris is the most massive currently known dwarf planet, with a satellite, Dysnomia \citep{BS07}. 
Recently \citet{Holler2021} obtained an updated orbit of Eris' satellite, Dysnomia, with a corrected orbital period of $P_{orb}$\,=\,15.785899$\pm$0.000050\,d. They suggested various possible reasons for the observed non-Keplerian orbit of the satellite, including the precession of Dysnomia's orbit due to the oblateness of Eris, an irregularly shaped Dysnomia, an unseen interior satellite, or center-of-light versus center-of-body offsets. 

Several light curve studies can be found in the literature providing very different rotation periods. 
\citet{Lin2007} obtained a light curve period of 3\fh55 with an amplitude of $\Delta$m\,$\leq$\,0\fm05 using a 1\,m-class ground based telescope.  
A low amplitude visual light curve of Eris was tentatively detected by 
\citet{Roe2008}, with a period of P\,=\,1\fd08$\pm$0\fd02, and 
with a peak-to-valley amplitude upper limit of $\sim$0\fm1 based
on Swift satellite data. They also reported that the shape of the light
curve is likely not sinusoidal, indicating the presence of a dark patch 
which is visible in part of the rotation period only. 
\citet{Duffard2008} obtained a light curve
period of 13\fh7 at a high confidence level, also using a 1\,m class
telescope. 
\citet{Carraro2006} did not obtain a definite rotation period, just a lower limit of $\sim$5\,d, and also \citet{Rabinowitz2007} and \citet{Sheppard2007} could not identify any period in their data. \citet{Rabinowitz2014DPS} reported on a possible 
synchronous rotation of the Eris-Dysnomia system, with a dominant periodicity in
the light curve matching the orbital period of $\sim$15.8 days. 
\citet{Holler2020} suggested that the rotation of Eris is near-synchronous, with a period of P\,=\,14.56$\pm$0.01\,d, indicating that the system is not yet fully tidally evolved. 
%

In this paper we present the analysis of long-term brightness monitoring data collected from various instruments, including the TESS and Gaia space telescopes, and several ground-based telescopes which covered different possible period ranges from a few hours to a rotation synchronized with the orbital motion of the satellite. Due to its sampling rate and duration TESS data could be used to investigate periods from a few hours to a few days. 
Ground based data were typically measured in blocks covering a few nights, with (very) long gaps between the blocks which allowed us to search for light curve periods in the few day -- synchronized rotation (15.8\,d) range. The sparse sampling of the Gaia data allowed us to search for rotation periods in this latter range, too. These observations and the data reduction are described in detail in Sect. \ref{sect:obs}. The summary of the analysis of these data and the description of curve period identification are presented in Sect.~\ref{sect:results}. Using the currently known characteristics of the system we applied a simple tidal evolution model to try to match the rotation period which we obtained from our light curve measurements (Sect. \ref{sect:tidal}). Our conclusions are given in Sect. \ref{sect:concl}.

\def\amin{\ifmmode ^{\prime}\else$^{\prime}$\fi}
\def\farcs{\hbox{$.\!\!^{\prime\prime}$}}  

\section{Periods identified in the different data sets}\label{sect:results}

\subsection{TESS \label{sect:tessresults}} 
TESS space telescope data could be used to search for possible light curve periods in the range of a few hours to a few days (see Sect.~\ref{sect:tess}), and we could identify a period with a residual minimum at a frequency of f\,=\,0.411$\pm$0.018\,cycle/day (denoted as c/d hereafter) which corresponds to a period of P\,=\,58.394$\pm$2.571\,h, with a light curve amplitude of $\Delta$m\,=\,0.132$\pm$0.037\,mag, after correcting for instrumental effects. This frequency, however, is considered to be tentative (1.8\,$\sigma$) due to the significantly increased noise at frequencies below 1\,c/d. At frequencies above 1\,c/d, however, a 1$\sigma$ upper limit of 0.03\,mag amplitude (peak-to-peak) could be obtained, indicating that no light curve period above this amplitude level is present at these shorter periods (P\,$\leq$\,24\,h). 

\subsection{Ground-based data \label{sect:groundbasedresults}}

We used a large set of ground based data (Sect.~\ref{sect:groundbased}), which are in part new measurements using several 1-2m-class telescopes (see Table~\ref{tab:phot_opt}), supplemented by measurements taken from the literature, including ground-based data from \citep{Carraro2006}, \citep{Rabinowitz2007}, \citep{Sheppard2007} and \citep{Duffard2008}, and the Swift satellite data from \citet{Roe2008}. We used a residual minimalization method (see Sect.~\ref{sect:periodfittingmethod}) to find the light curve amplitude and period best matched by this large data set. The efficiency of the method was tested using synthetic light curves, using a sampling similar to the real Eris data.  We assumed that i) the light curve amplitude is the same in any of the photometric bands used and ii) the light curve can be characterised by a simple sinusoidal variation. With these assumptions each model light curve can be described by four parameters: light curve amplitude, period, phase-shift, and an offset from the photometric zero point. We allowed a different zero-point offset for each measurement block (data consisting of measurements of consecutive nights) even if the data were taken by the same instrument and filter combination due to the sometimes year-long gaps between the measurement blocks. The best-fitting light curve period and amplitude is characterised by the minimum in the C(P,$\Delta m$) function, obtained by the residual minimalization, where P and $\Delta m$ are the period and amplitude of the light curve, respectively (see Sect.~\ref{sect:periodfittingmethod}).

\begin{figure}[ht!]
    \centering
   \includegraphics[width=0.47\textwidth]{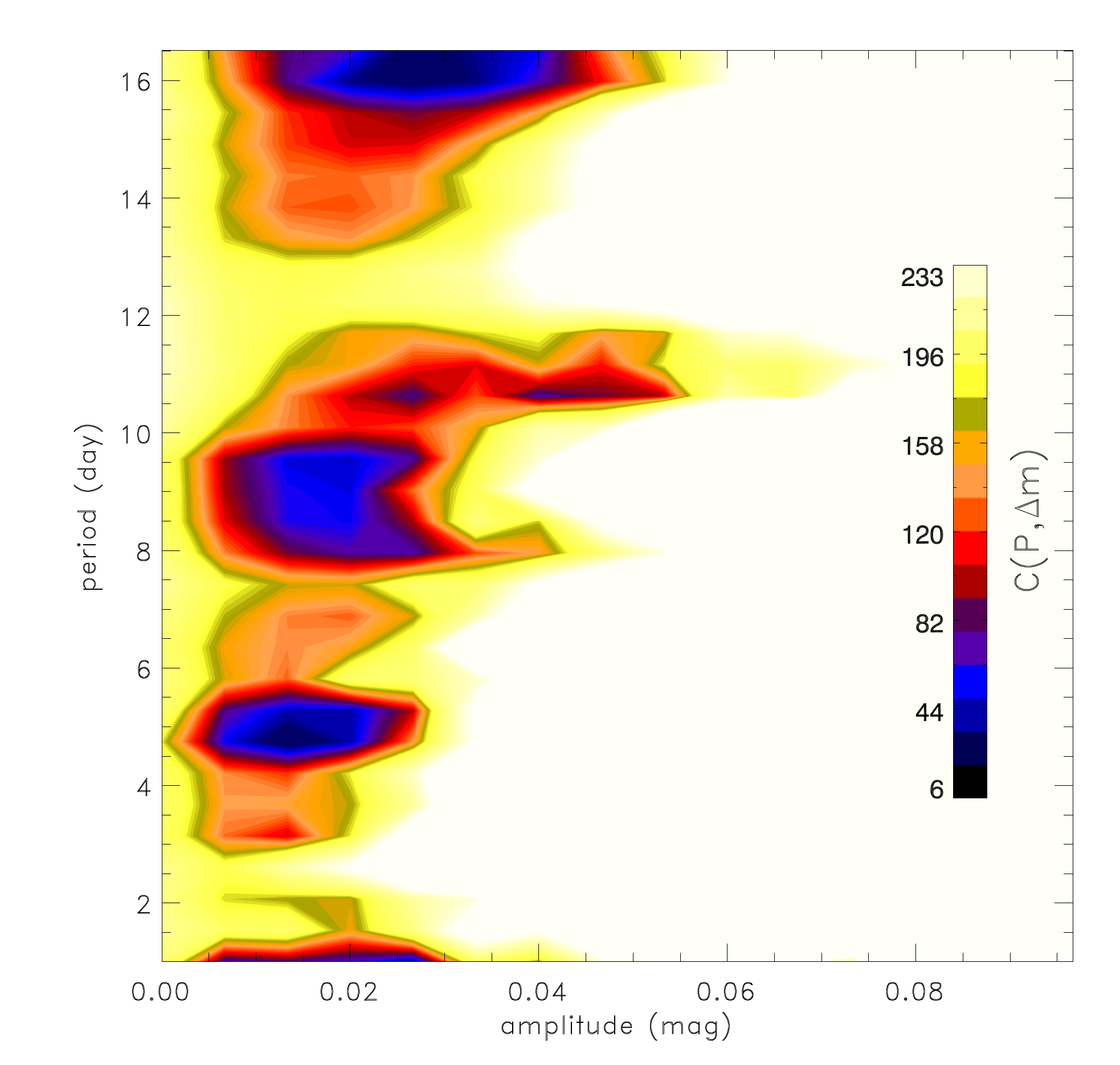}
    \caption{C(P,$\Delta m$) contour map; The most prominent minima is identified at a period of $\sim$16\,d, very close to the orbital period of Dysnomia, 15.78\,d (see the text for details).}
    \label{fig:chi2contour}
\end{figure}

The possible light curve periods were chosen in the range P\,$\in$\,[1,17]\,d. The upper limit was set to sufficiently cover the synchronized period (15.8\,d), while the lower limit was set to 1\,d as we used 'nightly' average values in many cases. 
We restricted our amplitude search range to  $\Delta m$\,$\in$\,[0,0.2]\,mag (peak-to-peak, i.e. twice the sine amplitude), 
as the original $\Delta m$\,$\in$\,[0,0.6]\,mag range was found to be too wide and did not provide minima in the large amplitude domain.
%
Our results are presented in Fig.~\ref{fig:chi2contour}. While the C(P,$\Delta m$) contour map shows several shallower minima, there is one main minimum, at P\,=\,16.2$\pm$0.5\,d, and $\Delta m$\,=\,0.027$\pm$0.005, very close to the orbital period of Dysnomia (15.8\,d). To check the robustness of this result, and obtain the period and amplitude uncertainties, we repeated the process by modifying the photometric data points by adding a random value with normal distribution using the specific measurement errors as standard deviations, and repeating the fitting process several times for the whole data set. 

\subsection{Gaia \label{sect:gaiaresults}}

\begin{figure}[ht!]
    \centering
    \includegraphics[width=0.5\textwidth]{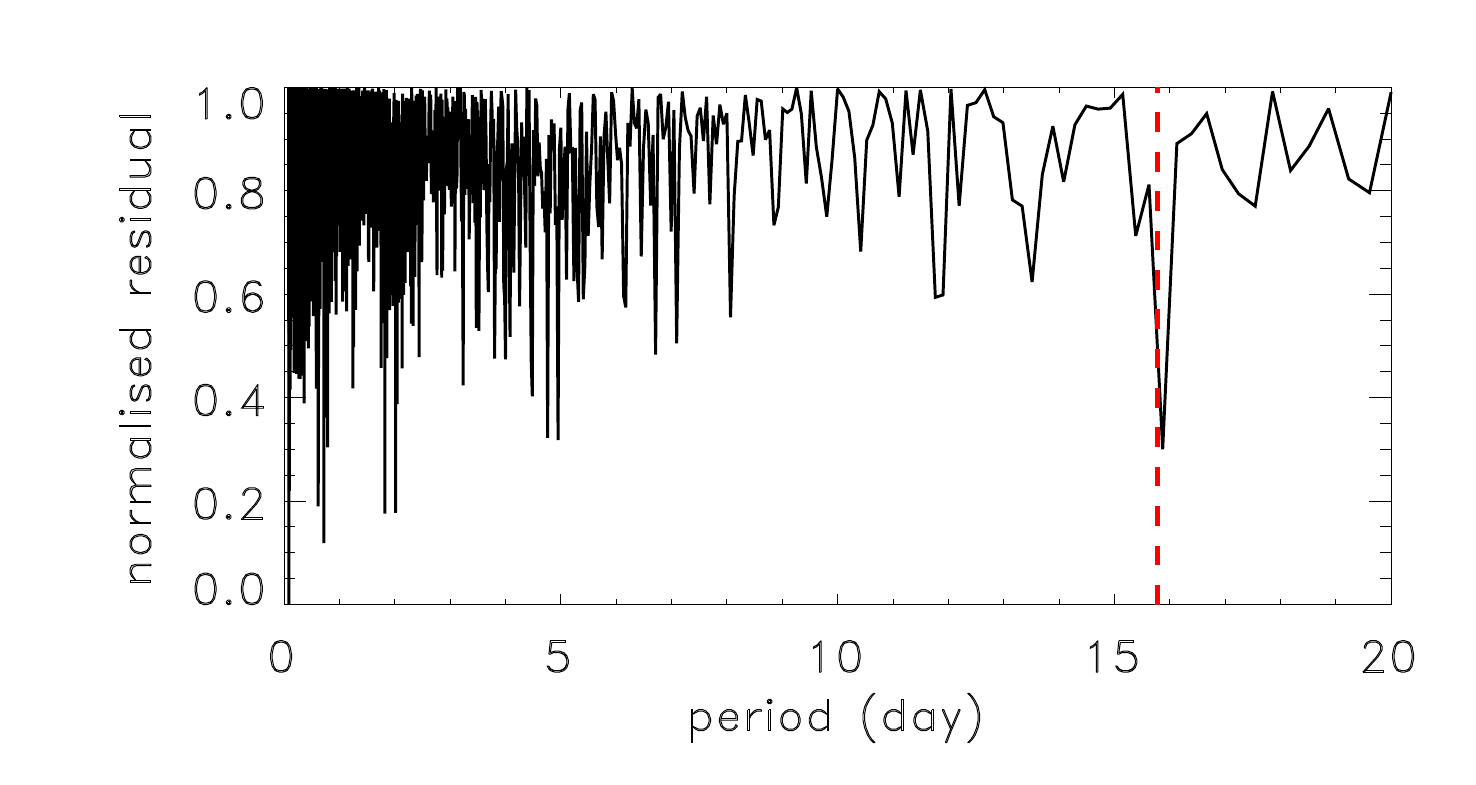}
    \includegraphics[width=0.5\textwidth]{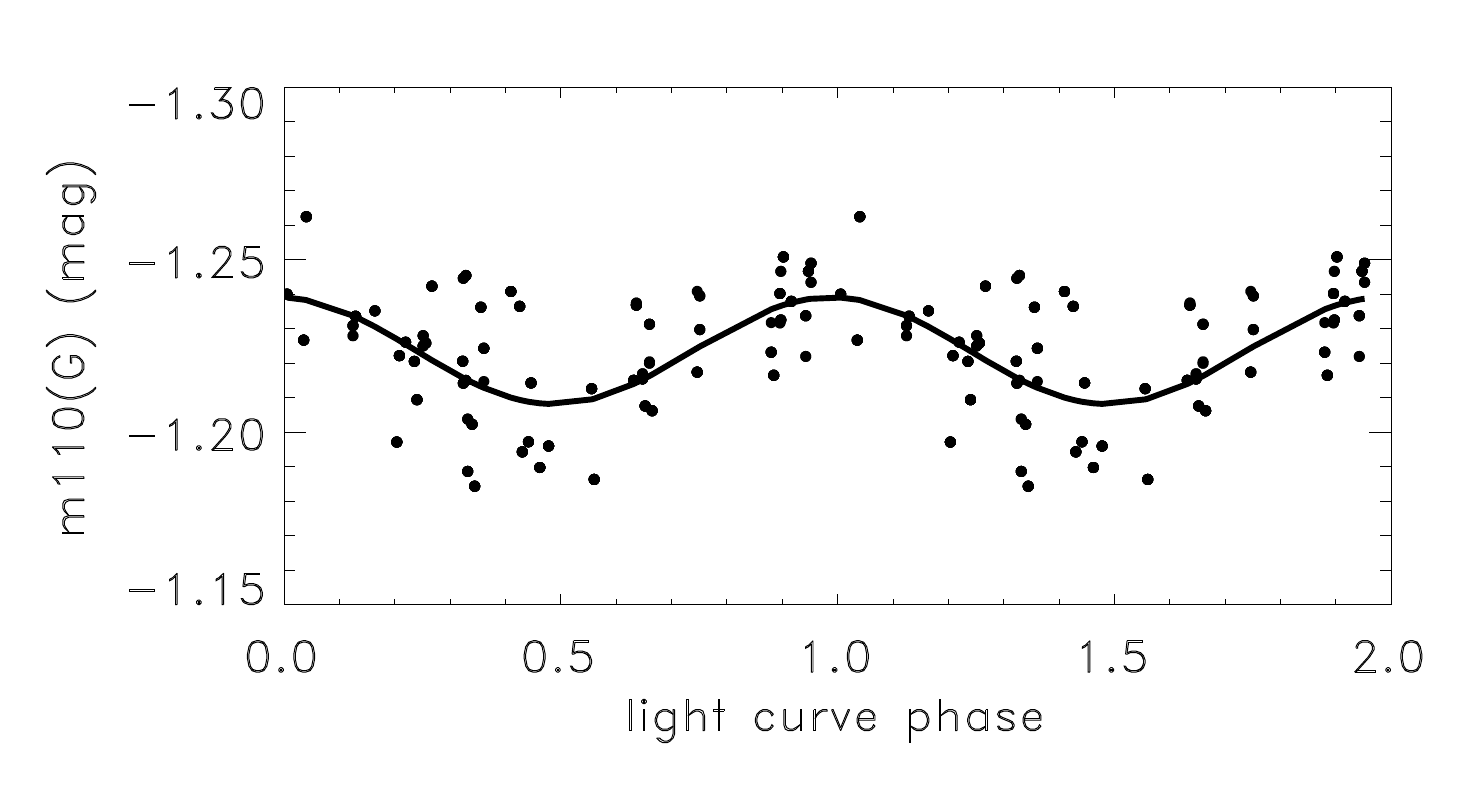}
    \caption{Upper panel: Normalized residual spectrum of the Eris Gaia light curve. The red dashed line is at P\,=\,15.78\,d, the orbital period of Dysnomia; Lower panel: Gaia light curve folded with f\,=\,0.063\,c/d. The solid curve is the best-fit sinusoidal, with a peak-to-peak amplitude of $\Delta m$\,=\,0.031$\pm$0.001.}
    \label{fig:gaialc}
\end{figure}

Gaia G-band photometry data of Eris (see Sect.~\ref{sect:Gaia_data}) was analysed using a residual minimalization algorithm to identify the possible frequencies in the light curve (see Fig.~\ref{fig:gaialc}, upper panel). There is one strong minimum identified at the long period part of the residual spectrum at 15.87$\pm$0.22\,d (5.5\,$\sigma$), very close to the 15.78\,d orbital period of Dysnomia. 
The Gaia light curve folded with this period (Fig.~\ref{fig:gaialc}, bottom panel) was fitted with a sinusoidal curve using a Levenberg-Marquardt fitter which provided a peak-to-peak amplitude of $\Delta$m\,=\,0.031$\pm$0.001\,mag. 

Both the ground based data (Sect.~\ref{sect:groundbased}) and the Gaia data strongly suggest that the light curve period of the Eris-Dysnomia system is equal to the orbital period of Dysnomia. 
As Dysnomia is significantly fainter than Eris in the visible range \citep[1:0.0021, see][]{BS07} a light curve variation of $\sim$0.03\,mag has to be associated with Eris, and not with Dysnomia. 
The P\,=\,15.78\,d period rules out all shorter periods in the few hours -- few days range (see a detailed list in Sect.~1), as well as the semi-synchronized rotation period of P\,=\,14.56\,d obtained by \citet{Holler2020}. We, however, cannot exclude that the rotation period is not exactly the orbital period, but very close to it. Considering the ground-based and Gaia data the latter one has a smaller uncertainty in the period determination, providing an uncertainty of $\sim$5\,h. As we show below, the system must be extremely fine-tuned to have an actual rotation period so close to but different from the synchronized case, assuming a simple binary system. Therefore we argue that the rotation of Eris and Dysnomia is double-synchronized, i.e. the Eris-Dysnomia system is fully tidally locked.   

As shown in Sect.~\ref{sect:grond} Eris appears to have a considerable variability in the near-infrared J-H colour, while all measurements show rather similar colours in the visible.  
One explanation for this behavior could be that the surface composition of Eris is not homogeneous and parts of the surface are covered with ices which have characteristic bands in the near-infrared, but have a reflectance similar to other materials in the visible \citep{FernandezValenzuela2021}. 
E.g. Eris is known to have strong methane features in its reflectance spectrum \citep{AlvarezCandal2020}, especially between 1.5-1.8 $\mu$m, close to the H-band. 
A variegation in surface composition may lead to a rotational variation in the near-infrared (J-H) colour, while leaving the visible range colours unchanged. 

\section{Possible tidal evolution scenarios}\label{sect:tidal}


We used a simple tidal evolution model (see Sect.~\ref{sect:tidalev}) to find the possible initial conditions and physical characteristics of Eris and Dysnomia that could have led to the currently observed tidally locked rotation of Eris. Our main assumption is that the Eris-Dysnomia system is formed in a giant collision and started tidal evolution from a much more compact configuration, Eris spinning significantly faster than today \citep{Ragozzine2009,Barr2016,Arakawa2019}. A specific model is characterised mainly, on one hand, by the properties of Eris relevant for tidal interactions, including the tidal dissipation factor $Q_p$, the rigidity $\mu_p$ and/or the second order tidal Love number $k_{2p}$; and by the mass $M_s$ and effective radius $R_s$ (or, equivalently the density $\rho_s$) of the satellite, Dysnomia. Previously \citet{GB2008} studied the tidal evolution of the Eris-Dysnomia system, but they restricted their calculations to a specific Dysnomia radius of $R_s$\,=\,75\,km, and mass of $M_s$\,=\,2.3$\cdot$10$^{18}$\,kg. In our model we considered a wide range of $R_s$ values which are compatible with the brightness constraints and allows radius and mass values as large as $R_s$\,$\approx$\,370\,km and $M_s$\,$\approx$\,5$\cdot$10$^{20}$\,kg (see Sect.~\ref{sect:tidalev}) 

\begin{figure}
\begin{center}
    \includegraphics[width=0.48\textwidth]{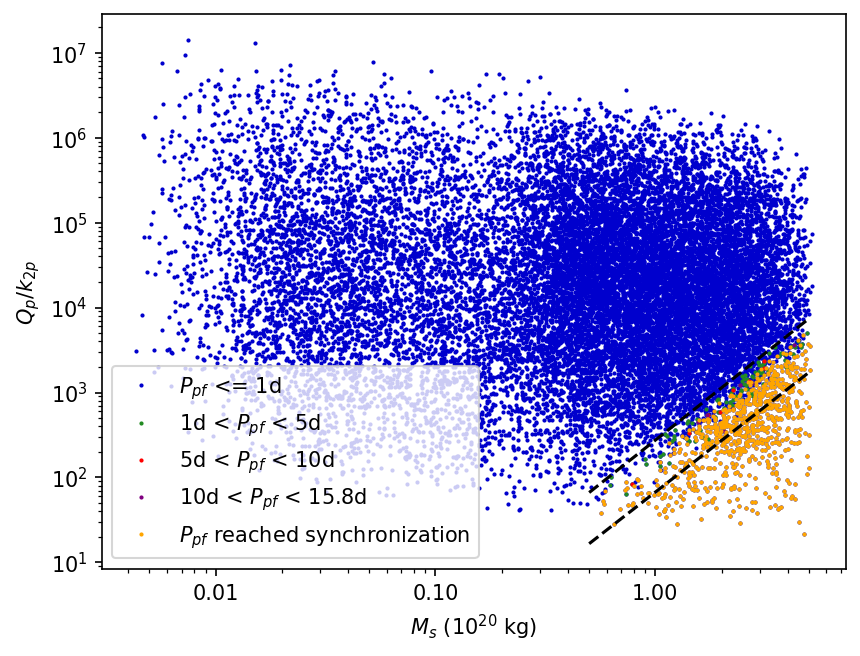}
    \caption{$Q_p/k_{2p}$ versus the mass of Dysnomia $M_s$ in the first trial runs. Orange symbols mark those cases which ended up in synchronized Eris rotation; blue symbols represent the cases with P\,$<$\,1\,d final Eris rotation periods. 
    Dashed lines mark the region of $M_s$--$Q_p/k_{2p}$ values that lead to slowed down rotation or tidal synchronization of Eris (see text for details.)}
    \label{fig:firstrun}
\end{center}
\end{figure}

We first run our code for a large set of models which covered a wide range of possible parameter values as shown in Fig.~\ref{fig:firstrun}. For most of these cases the final rotation periods of Eris, $P_{pf}$, remained below $P_{pf}$\,$\leq$\,1\,d, but there is a well defined area in the $Q_p/k_{2p}$--$M_s$ plane where $P_{pf}$\,$>$\,1\,d, or we reached synchronisation. 
This area is defined approximately defined by the two dashed lines in Fig.~\ref{fig:firstrun} which roughly satisfy a $Q_p M_s^2/k_{2p}$\,=\,constant relationship, as expected from the calculation of the spin rate change (Eq.~\ref{eq:omegadot}). As we are interested especially in those cases when the rotation of Eris slowed down considerably, we selected starting parameters from the area defined above on the $Q_p/k_{2p}$ vs. $M_s$ plot (Fig.~\ref{fig:firstrun}) to further map the parameter space in a second set of runs. The results of these runs are presented in Fig.~\ref{fig:simplots}. 

Even in our simple model we have a rather wide range of parameters which lead to synchronized or nearly-synchronized rotation rates for Eris. However, there are some general conclusions which can be drawn from our simulations. First, it is feasible that a massive Dysnomia can considerably slow down the rotation of Eris, even forcing it to a synchronized rotation state. E.g. Eris rotation periods of 10\,d\,$\leq$\,$P_{pf}$\,$\leq$15.78\,d can be reached for Dysnomia-to-Eris mass ratios of 10$^{-2}$\,$\leq$\,q\,$\leq$\,3$\cdot$10$^{-2}$. To reach such a large mass, Dysnomia has to be dark, with p$_{Vs}$\,$\lesssim$\,0.06, in contrast to the very bright surface of Eris ($p_{Vp}$\,=\,0.96). While it is possible to generate (near-)synchronous rotation for  p$_{Vs}$\,$\lesssim$\,0.06, most of these runs required p$_{Vs}$\,$\lesssim$\,0.04. Such a dark surface, and a corresponding large size is also suggested by the submm detection of Dysnomia with ALMA \citep{BB18}, implying $R_s$\,=\,350$\pm$58\,km and $p_V$\,=\,0.04$^{+0.02}_{-0.01}$.



\begin{figure}
    \centering
    \includegraphics[width=\columnwidth]{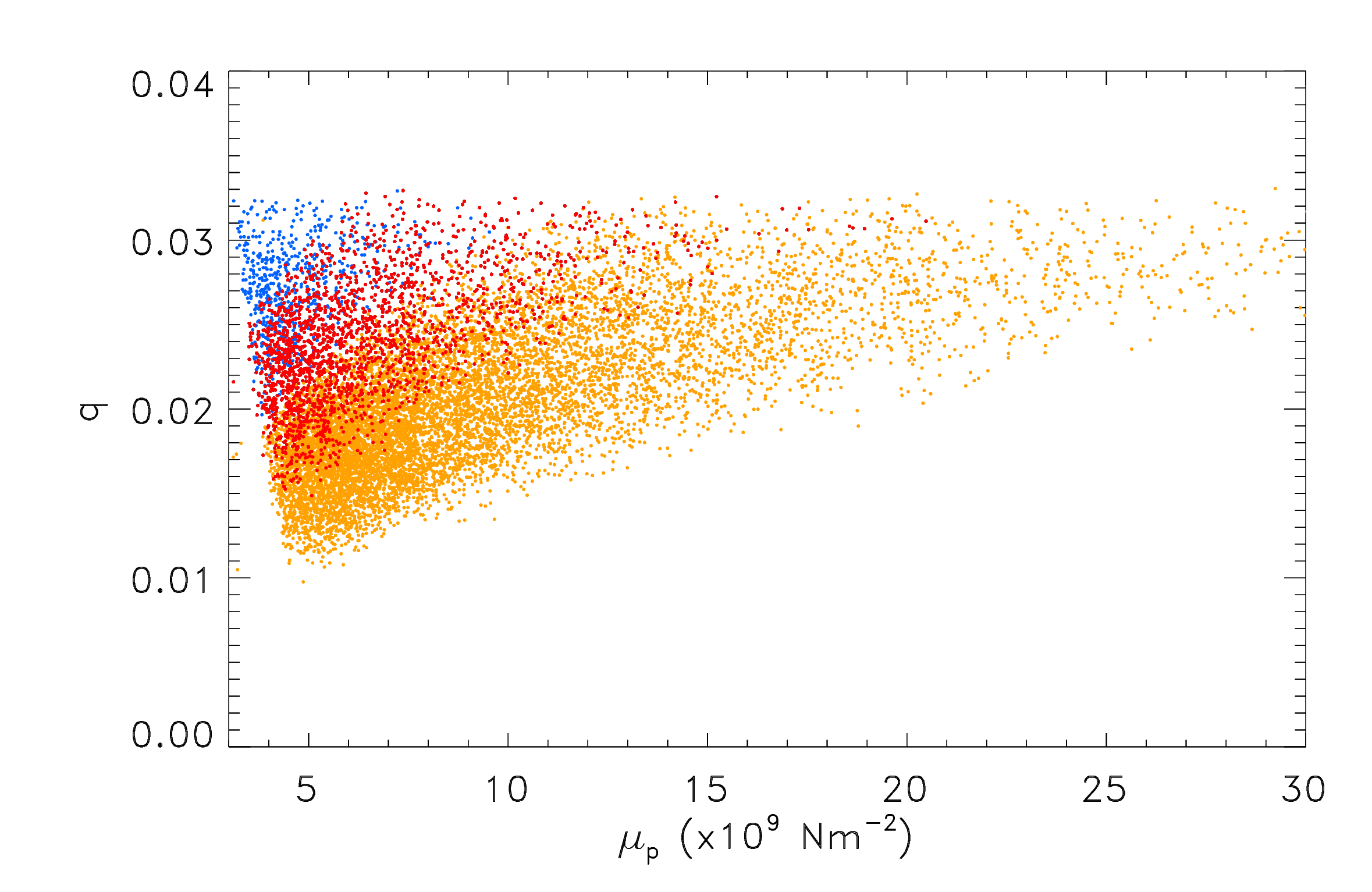}
    \caption{Dysnomia-to-Eris mass ratio (q) versus the rigidity of Eris ($\mu_p$) in those simulation runs when Eris ended up in a synchronized rotation. The orange, red and blue colours mark Eris tidal parameters of Q$_p$\,=\,50, 100 and 200 ($\pm$10\%).}
    \label{fig:q_mup_qp}
\end{figure}

Although our simulations are run for a wide range of material/tidal parameters assumed for Eris, mainly $Q_p$ and $\mu_p$, these parameters are canonically chosen in a much narrower range. In Fig.~\ref{fig:q_mup_qp} we selected those simulation runs for which the tidal quality parameters of Eris was Q\,=\,50, 100 or 200 ($\pm$10\%) -- $Q_p$\,=\,100 is the canonical value usually assumed in the TNO tidal evolution calculations -- and Eris' rotation became tidally locked. For $Q_p$\,=\,100 (red symbols in Fig.~\ref{fig:q_mup_qp}) synchronization is reached for Dysnomia-to-Eris mass ratios of $q$\,=\,0.02-0.03, depending on the Eris rigidity parameter $\mu_p$ which, for this $Q_p$ value, can be in the range 4-20$\cdot$10$^9$\,N\,m$^{-2}$. These $\mu_p$ values correspond to the rigidity of ice ($\sim$4$\cdot$10$^9$\,N\,$^{-2}$), or a mixture of 'ice and rock', in the case of the higher $\mu_p$ values \citep[c.f.][]{Grundy2011}. To obtain mass ratios of $q$\,=\,0.02-0.03 Dysnomia has to be large (D\,$\gtrsim$\,600\,km) and its density has to be in the range $\rho_s$\,=\,1.8-2.4\gcc. These cases are also associated with very low, $p_V$\,=\,0.02-0.03 geometric albedos. In the case of a higher tidal quality parameter value, e.g. $Q_p$\,=\,200, even higher mass ratios are required, and correspondingly the Dysnomia densities are also higher, $\rho_s$\,$\geq$\,2.0\gcc. In these cases the allowed rigidity of Eris is 3$\cdot$10$^{9}$\,$\lesssim$\,$\mu_p$\,$\lesssim$\,8$\cdot$10$^{9}$\,N\,m$^{-2}$. A lower $Q_p$, however, would allow smaller Dysnomia masses from q\,$\approx$\,0.01, with a significant dependence on the rigidity as higher q is required for higher rigidity values, up to $\mu_p$\,=\,3$\cdot$10$^{10}$\,N\,m$^{-2}$. Due to the lower required mass the density range allowed for Dysnomia is also wider, $\rho_s$\,=\,1.2-2.4\gcc. As indicated by the full range of simulations (see Fig.~\ref{fig:simplots}) progressively smaller values of $Q_p$ will allow smaller Dysnomia mass values to be compatible with a synchronized Eris rotation. At $Q_p$\,=\,10, our smallest value chosen, a Dysnomia with a mass ratio of q\,$\approx$\,0.006 and density of $\rho_s$\,$\approx$\,1.0\,\gcc\, would be massive enough to lock the rotation of Eris. 
Precise absolute astrometry of the primary and secondary which could be performed with ALMA \citep{BB19} may detect the barycentric wobble and obtain the mass ratio, also putting constraints on the tidal quality factor $Q_p$.
 
\section{Conclusions}\label{sect:concl}

In this paper we analysed long-term ground based photometric observations of Eris, complemented by measurements with the TESS and Gaia space telescopes. While the TESS data did not provide a conclusive rotation period, both the combined ground-based measurements and the Gaia data unambiguously point to a light curve period that equals to the orbital period of Dysnomia, i.e. a tidally locked rotation of Eris. The synchronized rotation of Eris -- which is considered to be the consequence of a tidal interaction with Eris' moon Dysnomia -- puts constraints on the key physical properties of the satellite, as well as on those of Eris, as discussed in Sect.~\ref{sect:tidalev}. While the light curve or tidal evolution results does not directly constrain the shape, due to its very low spin rate the shape of Eris is expected to be very close to spherical, similar to that observed for Pluto and Charon \citep{Nimmo2017}. For Eris, both a homogeneous density interior model with a Maclaurin shape, or the Darwin-Radau model \citep[see e.g.][]{MD2000} with a two-component, rocky core and an ice mantle interior provides flattening values of  $\epsilon$\,$\leq$\,0.0001. This also means that the occultation shape and size solution \citep{Sicardy2011} has to be very close to the spherical one with R\,=\,1163$\pm$6\,km, an important constraint e.g. for thermal emission models. Also, \citep{Holler2021} suggested an oblate Eris as a possible cause of the non-Keplerian orbit of Dysnomia which now seems to be a less likely option.


Another explanation of the non-Keplerian Dysnomia orbit could be a centre of light -- centre of body (CoL-CoB) offset due to a large albedo pattern on the surface or Eris. Considering our best fitting $\Delta m$\,=\,0.031\,mag amplitude we investigated this scenario by assuming a single spot on the equator, coincident with the orbital plane of Dysnomia, and a viewing geometry as defined by the 'combined' solution in \citep{Holler2021}. In our simple model the spot is visible in a fraction of the rotational phases and completely invisible in others; in these latter cases Eris has a homogeneous, high albedo surface. We varied the size of the spot with the spot albedo in a way that it always produces the required light curve amplitude, and we considered both Lommel-Seeliger and Lambert scattering laws. With this light curve amplitude the maximum CoL-CoB offset that could be obtained is $\sim$40\,km both for the Lommel-Seeliger and the Lambert scattering and roughly similar values are obtained for the whole range of albedos considered. This is much smaller than the 462\,km offset obtained by \citet{Holler2021}. Considering the maximum possible size of $\sim$600\,km for Dysnomia, it is not feasible that the dominant part of the CoL-CoB offset could be due to features on the surface of the satellite, i.e. CoL-CoB offset is not a likely reason for the non-Keplerian orbit. 
The non-Keplerian orbit of Dysnomia could also be caused by a non-spherical shape of the satellite. A recent study of Kepler/K2 light curves of transneptunian objects \citep{kecskemethy2022} shows that light curve amplitudes of TNOs remain larger at large (D\,$\gtrsim$\,400\,km) sizes where the asphericity of main belt asteroids drops significantly \citep{Vernazza2021}. While this could be due to an irregular-to-spherical transition at larger sizes for TNOs, their general low densities and high porosities point against this scenario. At the expected sizes of Dysnomia (D\,$\gtrsim$\,600\,km) objects should be fairly round, even with higher densities and a considerable internal strength. 

 
Assuming that the Eris-Dysnomia system formed in a giant impact the rotation period of the post-impact Eris was probably much shorter, on the order of a few hours. This fast rotation had to be slowed down by the tidal interaction with Dysnomia. As we have shown above, to reach synchronized rotation periods Dysnomia has to be relatively massive (mass ratio of q\,=\,0.02-0.03), assuming canonical values for the $Q_p$ tidal dissipation factor and $\mu_p$ rigidity of Eris.
This mass ratio is the second largest value in the transneptunian region after the $\sim$8:1 ratio in the Pluto-Charon system \citep[see][for earlier evaluations]{Barr2016,Arakawa2019,Kiss2019}. (We note that currently the mass ratio in the Orcus-Vanth system is rather uncertain). The relatively high mass ratio is also associated with high Dysnomia densities of $\rho_s$\,=\,1.8-2.4\,\gcc\, which are much larger than the typical densities of transneptunian objects in this size range, $\rho$\,=\,0.5-1.0\,\gcc\, \citep[see e.g.][]{BN19}. In our tidal evolution model unconventionally low $Q_p$ tidal dissipation factors would allow lower Dysnomia densities (down to $\rho_s$\,$\approx$\,1.2\gcc) and Eris reaching synchronized rotation at the same time, however, these values are still above the typical low densities of R\,$\approx$\,300\,km objects, and would also require a low level of porosity. Collisional simulations have shown that in general intact moons with 10$^{-3}$\,$\leq$\,q\,$\leq$\,10$^{-1}$ could form in the transneptunian region assuming a wide range of impact parameters \citep{Arakawa2019}. More detailed impact and tidal evolution simulations should be able to identify the conditions which could lead to the present high density Eris -- high density Dysnomia system. 



\begin{acknowledgements}

The research leading to these results has received funding 
from the K-138962 grant of the National Research, Development and Innovation Office (NKFIH, Hungary). The data presented in this paper were obtained from the Mikulski Archive for Space Telescopes (MAST). STScI is operated by the Association of Universities for Research in Astronomy, Inc., under NASA contract NAS5-26555. Support for MAST for non-HST data is provided by the NASA Office of Space Science via grant NNX09AF08G and by other grants and contracts. This research has made use of data and services provided by the International Astronomical Union's Minor Planet Center. Part of the funding for GROND (both hardware as well as personnel) was generously granted from the Leibniz-Prize to Prof. G. Hasinger (DFG grant HA 1850/28-1). We are grateful to the CAHA and OSN staff. This research is partially based on observations collected at the Centro Astronómico Hispano Alemán (CAHA) at Calar Alto, operated jointly by Junta de Andalucía and Consejo Superior de Investigaciones Científicas (IAA-CSIC). This research was also partially based on observation carried out at the Observatorio de Sierra Nevada (OSN) operated by Instituto de Astrofísica de Andalucía (CSIC). P.S-S. acknowledges financial support by the Spanish grant AYA-RTI2018-098657-J-I00 “LEO-SBNAF”(MCIU/AEI/FEDER, UE). P.S-S., J.L.O., N.M., and R.D. acknowledge financial support from the State
Agency for Research of the Spanish MCIU through the “Center of Excellence Severo Ochoa” award for the Instituto de Astrofísica de Andalucía (SEV-2017-0709), they also acknowledge the financial support by the Spanish grants AYA-2017-84637-R and PID2020-112789GB-I00, and the Proyectos de Excelencia de la Junta de Andalucía 2012-FQM1776 and PY20-01309. We are also thankful to our reviewer for the useful comments. This work made use of Astropy:\footnote{\url{http://www.astropy.org}} a community-developed core Python package and an ecosystem of tools and resources for astronomy \citep{astropy:2013, astropy:2018, astropy:2022}. 
\end{acknowledgements}





\scriptsize
\setlength{\bibsep}{0pt}
\bibliographystyle{aa_url}
\bibliography{tno.bib}

\begin{appendix}

\section{Supporting Material}

\subsection{Observations and data reduction}
\label{sect:obs}

\subsubsection{TESS data}
\label{sect:tess}
\normalsize

\begin{figure}[ht!]
    \centering
    \includegraphics[width=0.5\textwidth]{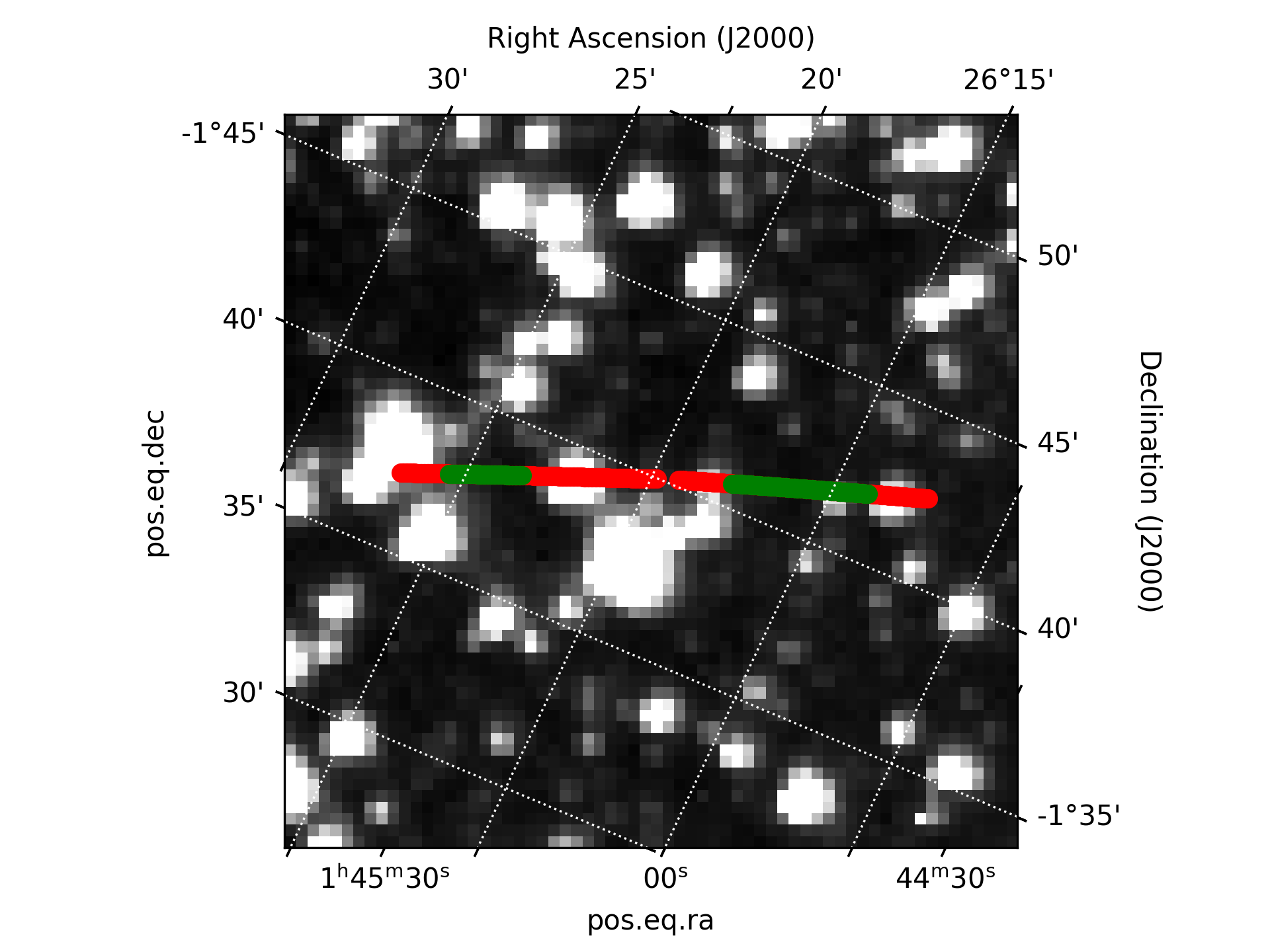}
    \caption{Eris' path through the field-of-view of Sector 3 / Camera 1 / CCD 3 of the TESS space telescope between Julian dates 2459115.89 and 2459142.52. The red and green parts mark those points that were excluded / considered for light curve analysis.}
    \label{fig:tessfov}
\end{figure}

The Transiting Exoplanet Survey Satellite \citep[TESS,][]{Ricker2015} observed Eris in Sector 30 with its Camera 1 and CCD 3 (Fig.~\ref{fig:tessfov}). 
The reduction of the TESS data was performed in the same way as described in \citet{Pal2020} which contains a detailed description of the reduction steps, photometry, and the derivation of the residual spectrum for frequency analysis. We only mention those steps here that are different from those in the \citet{Pal2020} pipeline. The TESS photometry data of Eris is provided in Table \ref{tab:phot_tess}.

\begin{table}[ht!]
    \caption{TESS photometry data of $(136199)$ Eris (sample).}
    \begin{center}
    \begin{tabular}{ccc}
        \hline
        Julian Date & $m$ & $\delta m$   \\ 
                    & (mag)    & (mag) \\
        \hline
        2459119.81621 & 18.99754 & 0.46354 \\
        2459119.85788 & 18.65064 & 0.33787 \\
        2459119.89954 & 18.69310 & 0.31456 \\
        ... & ... & ... \\
        \hline
    \end{tabular}
    \end{center}
    \footnotesize
    {\bf Note.} We are listing here the Julian date, brightness (m) and its uncertainty ($\delta$m) in the TESS photometric band. The table is available in its entirety in electronically readable format. All data points have 60\,min integration time. 
    \label{tab:phot_tess}
\end{table}

\begin{figure}[ht!]
    \centering
    \includegraphics[width=0.45\textwidth]{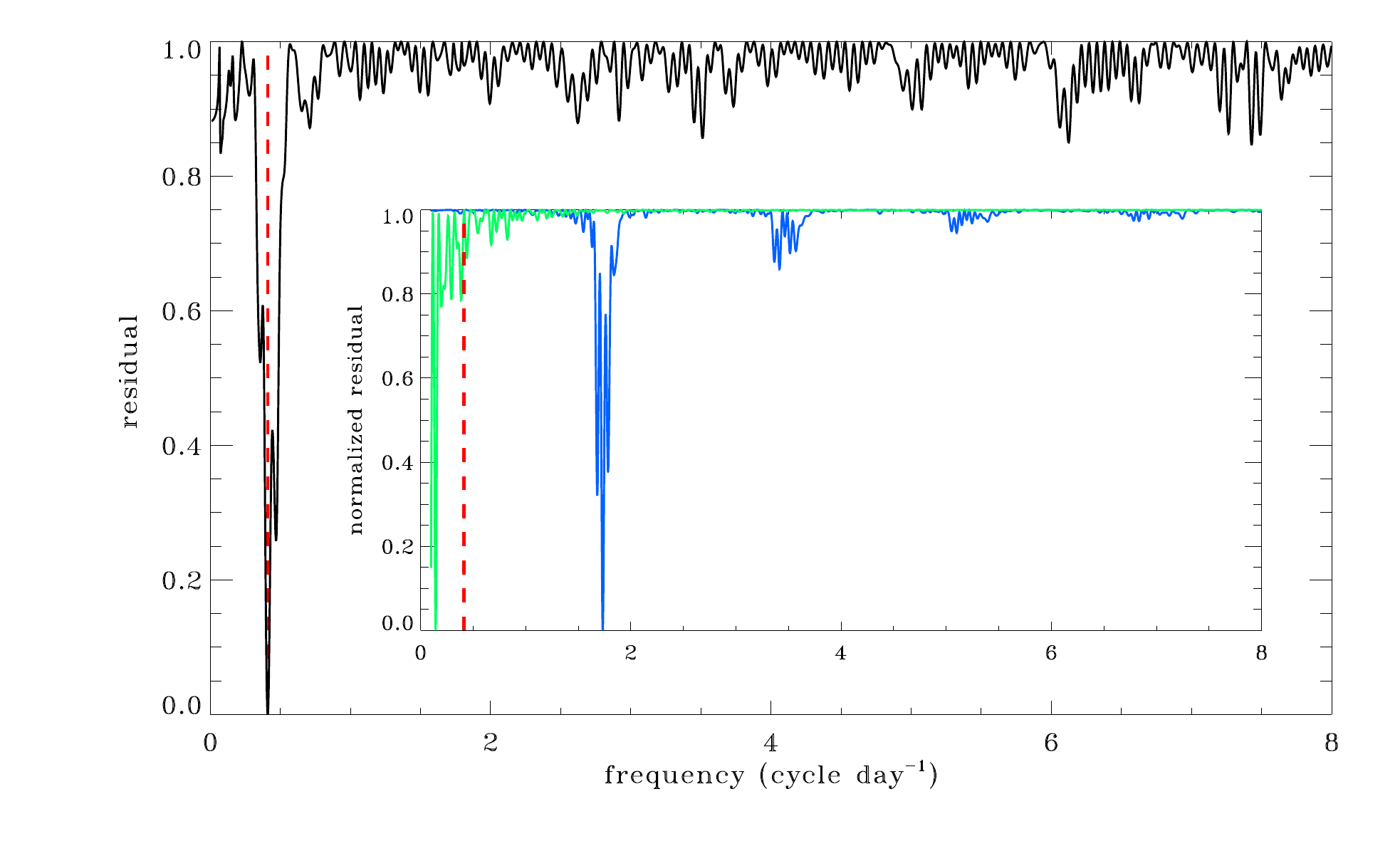}
    \includegraphics[width=0.45\textwidth]{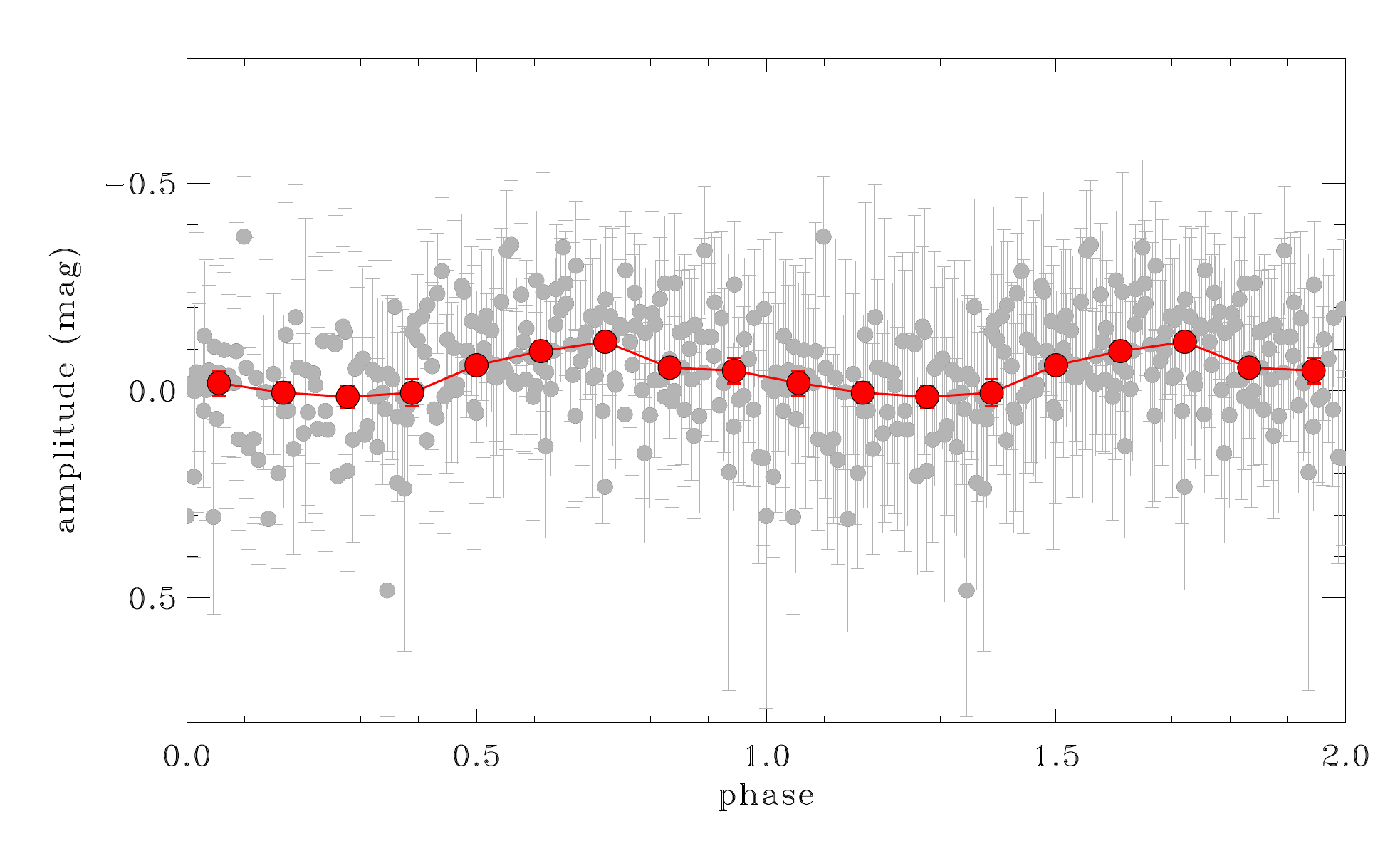}
    \caption{Top: Normalized residual spectrum of the TESS light curve of Eris. The insert shows the residual spectrum of the pixel-wise x- (blue) and y-direction (green) subpixel centroid positions. The most prominent characteristic frequency of f\,=\,0.411\,c/d is marked by a red vertical dashed line both in the main figure and the insert. Bottom: TESS light curve of Eris folded with f\,=\,0.411\,c/d. The red dots mark the binned light curve. These light curve data are presented in Table~\ref{tab:phot_tess}) in electronically readable format.}
    \label{fig:freq}
\end{figure}


\begin{figure}[ht!]
    \centering
    \includegraphics[width=0.5\textwidth]{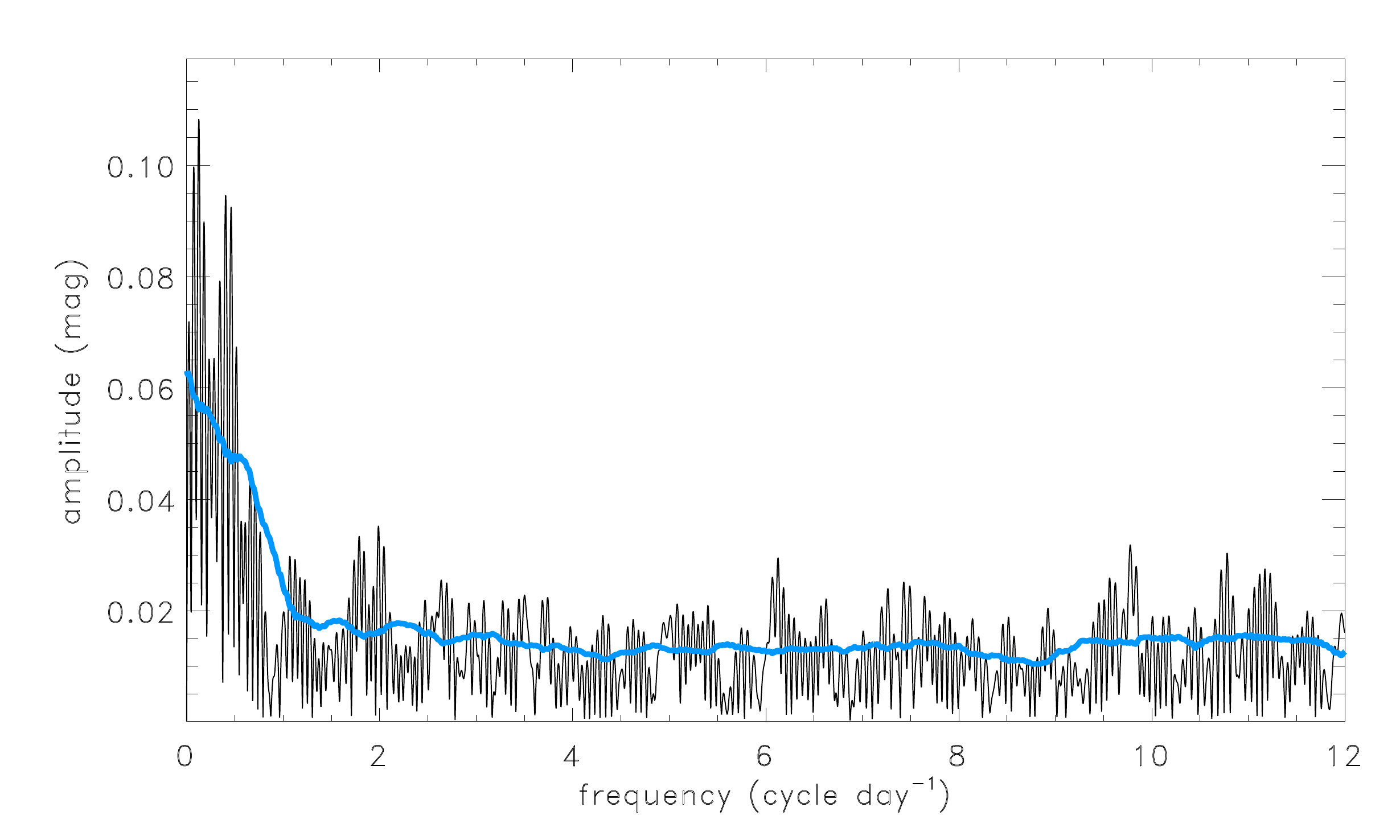}
    \caption{Fourier spectrum of the Eris TESS light curve. The blue curve is the r.m.s. amplitude, calculated using a running box and sigma clipping. }
    \label{fig:fourier}
\end{figure}

\normalsize 


A significant portion of the light curve data had to be excluded due to Eris' encounter with nearby background sources which left a dominant feature in the background-subtracted image. As shown in Fig.~\ref{fig:tessfov} the two 'green' zones, where the vicinity of Eris was relatively clean, covers two blocks with lengths of 2.3\,d and 6.6\,d. In these blocks readout-to-readout variation of the background was estimated to be $\sim$3\,mag lower then the typical ($\sim$18.5\,mag in the TESS bandpass) brightness of Eris, in the same measuring aperture. The residual spectrum obtained from the data of these two blocks merged is shown in Fig.~\ref{fig:freq}.

As TESS has large, 21\arcsec-sized pixels, the photometry of the source is affected by the relative position of the source inside the actual pixel, and the projection of the total source flux into the neighbouring pixels. This is expected to introduce a periodic signal as the target moves through the field-of-view, and the characteristic frequency depends on the actual apparent speed of the target in the X and Y (pixel-wise) directions. To look for this effect we checked the spectrum of the X and Y pixel fractions of centroid positions of Eris' TESS photometry. The results are presented in the insert in Fig.~\ref{fig:freq}. These residual spectra show well defined minima at f$_x$\,=\,1.73\,c/d (X-direction, blue curve in Fig.~\ref{fig:freq}) and f$_x$\,=\,0.14\,c/d (Y-direction, green curve). The residual spectrum of the TESS photometry data (main figure, black curve) shows a well-defined minimum at f\,=\,0.411$\pm$0.018\,c/d (P\,=\,58.394$\pm$2.571\,h). The uncertainty of the frequency is obtained as the FWHM of Gaussian fitted to the main frequency minimum in the residual spectrum. This frequency is different from the f$_x$ and f$_y$ frequencies identified above, and could not be associated with any currently known instrumental effects. The light curve folded with this frequency is shown in Fig.~\ref{fig:freq}. The peak-to-peak amplitude of this folded light curve is $\Delta m$\,=\,0.132$\pm$0.037\,mag, obtained the maximum-minus-minimum of the binned light curve as amplitude, and using standard error propagation in the calculation of the uncertainty.  

We also derived the Fourier spectrum of the Eris TESS light curve (Fig.~\ref{fig:fourier}, note the factor of two conversion between the Fourier and peak-to-peak amplitudes). 
The r.m.s. Fourier amplitude (blue) curve shows that the 1\,$\sigma$ noise is $\sigma_f$\,$\approx$\,0.015\,mag in the frequency range 1--12\,c/d, and it increases considerably for lower frequencies, reaching $\sim$0.065\,mag at the lowest ones. These r.m.s. amplitudes can be used to estimate light curve detection upper limits ($\Delta m_{lim}$) for specific frequencies. Considering the detection limit as 3\,$\sigma_f$ we obtain $\Delta m_{lim}$\,=\,0.045\,mag in the f\,=\,1--12\,c/d (P\,=\,2--24\,h) range, and $\Delta m_{lim}$\,=\,0.155\,mag at f\,$\approx$\,0.4\,c/d, i.e. at the frequency where a prominent peak was identified in the residual spectrum, as discussed above. While a peak in the Fourier spectrum can be identified at the same frequency (f\,=\,0.411\,c/d) as in the residual spectrum, the signal-to-noise value is significantly lower here, $\sim$1.8\,$\sigma$. Due to this limitation we consider the f\,=\,0.411\,c/d (P\,=\,58.394\,h) peak as tentative. Due to the limited length of the TESS light curve blocks considered in the analysis (2.3\,d and 6.6\,d) it was not possible to detect light curve periods longer than $\sim$3\,d, also excluding the possibility to detect periods close to the orbital period (15.78\,d). However, the Fourier spectrum shows that we can \emph{exclude} rotation periods in the range P\,=\,2--24\,h which would be associated with peak-to-peak amplitudes $\Delta m$\,$\geq$\,0.09\,mag, with a 3\,$\sigma$ confidence, i.e. if Eris had a rotation period in this range, it would have a very small amplitude, likely in the order of 0.03\,mag, or below. 

\subsection{GROND observations}
\label{sect:grond}

Observations of Eris with the GROND instrument \citep{gbc08} on the MPG 2.2m telescope at La Silla were made in 3 nights, with details given
in Tab. \ref{tab:other_obs}. Observations consisted of 8m4td observation blocks, i.e. individual 120\,s (115\,s) exposures in JHK (g'r'i'z') at each of the four telescope
dither (td) positions, except for the first epoch with 2\,s (20\,s)
exposures each at 2 td-positions. Since the Eris motion on the sky is below 0\farcs2/10 min, the 4 different dither pointings were co-added.
GROND data were  reduced in the standard manner \citep{kkg08}
using pyraf/IRAF \citep{Tody1993, kkg08}.
The optical imaging data ($g^\prime r^\prime i^\prime z^\prime$)
was calibrated against the Sloan Digital Sky Survey
(SDSS)\footnote{\url{http://www.sdss.org}} catalogue \citep{Eisenstein+2011},
and the NIR data ($JHK_{\rm s}$) against the 2MASS catalogue \citep{Skrutskie+2006}.
This results in typical 
absolute accuracies  of $\pm$0.03~mag in $g^\prime r^\prime i^\prime 
z^\prime$ and $\pm$0.05~mag in $JHK_{\rm s}$. 
Since the GROND dichroics were built after the Sloan filter system
\citep{gbc08}, the colour
terms are very small, below 0.01 mag, except for the $i^\prime$ band
which is substantially narrower than the SDSS i' band:
$i^\prime_{SDSS} - i^\prime_{GROND} = (-0.023\pm0.010) + (0.216\pm0.054) \cdot (i^\prime_{SDSS} - z^\prime_{SDSS})$
\footnote{\url{https://www.mpe.mpg.de/$~$jcg/GROND/calibration.html}}.
In order to minimize the impact of variability in the comparison stars, relative
photometry was done against the same observation (2010-08-31T08:51-09:03).
Since Eris moved by about 1\amin\ over the 3-day observing period, care was
taken to select comparison stars such that they were covered in all
observations. GROND photometry data are listed in Table~\ref{tab:phot_opt}. 
The GROND J and H colours in Table~\ref{tab:phot_opt} and the J-H colour index in Table~\ref{tab:grondcolours} are given in the AB photometric system. To convert them to the Vega system we used the conversions $J_{Vega} = J_{AB} - 0.91$ 
\label{eq:jab}
and $H_{Vega} = H_{AB} - 1.38$
\label{eq:hab}
from \citep{Blanton2007}. 

\begin{table}[ht!]
\caption{Ground-based photometry data of $(136199)$ Eris (sample).}
    \begin{center}
    \scriptsize
    \begin{tabular}{ccccccc}
        \hline
        Telescope & Julian Date & m & $\delta$m & Filt. & t$_{exp}$ & t$_{tot}$ \\
                  &             & (mag) & (mag) &       & (min)    & (min) \\
        \hline
        GROND & 2455436.91528 & 19.031 & 0.031 & $g^\prime$& 8 & 8\\
        GROND & 2455438.69348 & 19.097 & 0.010 & $g^\prime$& 8 & 8\\
        GROND & 2455438.75578 & 19.099 & 0.010 & $g^\prime$& 8 & 8\\
        \multicolumn{7}{c}{...} \\
        1.5m  & 2453647.53342 & 18.450 & 0.035 & R  & 293.3 & 450.1\\
        1.5m  & 2453648.55623 & 18.459 & 0.025 & R  & 353.3 & 393.0\\
        1.5m  & 2453649.56174 & 18.430 & 0.025 & R  & 383.3 & 428.8\\
        \multicolumn{7}{c}{...} \\
        \hline
    \end{tabular}
    \end{center}
    \footnotesize
    {\bf Note.} The table is available in its entirety in electronically readable format. The columns of the table are: Telescope name; mean Julian day of the measurement; target brightness (m) and its uncertainty ($\delta$m) in the respective filter band; filter; t$_{exp}$: sum of the individual exposure times used for this photometric point (as explained in the text); t$_{tot}$: difference between the end of the last, and the start of the first measurement used to obtain this specific photometric point. 
    \label{tab:phot_opt}
\end{table}

We compared the colours from our GROND photometry with values from the literature, as shown in Table \ref{tab:grondcolours}. We converted the GROND g,r,i,z colours to Johnson/Cousins BVRI, as described in \citet{Smith2002}. Near-infrared colours were converted from the respective systems used in the specific papers to a common 2MASS system, as described in the table caption. 
The colours from the new GROND measurements are in a relatively good agreement with the values from other studies in the visible bands. However, in the near-infrared, the J-H colours show large variations. J-H values range from -0.290$\pm$0.045 to 0.287$\pm$0.114, i.e. a $\sim$0.58 mag difference between the lowest and highest colour values.

\subsection{Other ground based observations \label{sect:groundbased}}
  
\begin{table*}
\caption{Summary table of ground-based observations.}
  \scriptsize 
   \begin{center}
     \begin{tabular}{cccccc}
    
     \hline
     \noalign{\smallskip}
     Telescope & Instrument & Date range & Number of nights & Filter & Reference \\
     \noalign{\smallskip}
     \hline
     \noalign{\smallskip}
     TESS & Camera 1 & 2020.09.27.-2020.10.16.& 19 & TESS Bandpass &\citep{Ricker2015}\\
     \noalign{\smallskip}
     MPG 2.2m & GROND &2010.08.28.-2010.08.31. & 3 & $g^\prime r^\prime i^\prime z^\prime$JH*&\citep{gbc08}\\
     \noalign{\smallskip}
     1.5m & Roper, Andor & 2005.10.03.-2020.10.16. & 18 & Johnson R, Clear &\tablefootnote{\url{https://www.osn.iaa.csic.es/en/page/15-m-telescope}}\\
     \noalign{\smallskip}
     CA2.2 & CAFOS2.2 & 2007.01.11.-2007.01.16. & 6 & Johnson R & \tablefootnote{\url{http://www.caha.es/CAHA/Instruments/CAFOS/}}\\
     \noalign{\smallskip}
     La Hita & SBIG STX-16803-3& 2014.10.24.-2014.10.28.& 5& Clear &\tablefootnote{\url{https://fundacionastrohita.org/instrumental/}}\\
     \noalign{\smallskip}
     \hline
     \end{tabular}
    \end{center}
    \label{tab:other_obs}
    \footnotesize
    {\bf Note.}  We are listing here the telescope, instrument (camera), date range, number of nights, and the filers used. *:GROND JH magnitudes are in the AB system. These observations, taken between JD\,=\, 2453647.53342 and 2459139.5323, covered heliocentric and observer distances of $r_h$\,=\,95.9-96.9\,au, $\Delta$\,=\,95.0-96.6\,au, and phase angles $\alpha$\,=\,0.12-0.58\,deg.
 \end{table*}    

We obtained ground based photometry data of Eris from four telescopes (see Table~\ref{tab:other_obs}).
All the data were reduced using standard calibration steps with the FITSH \citep{Pal2012} software package, i.e. bias, dark and flat corrections were applied. Then we performed aperture photometry on Eris and on selected comparison stars. Using the comparison stars as standard stars we obtained magnitudes from the Pan-STARRS DR2 catalogue and used these magnitudes to do an approximate standard calibration of Eris via simple linear fitting the observed and the catalogue magnitudes. In most cases only one filter was used, so more sophisticated standard calibration was not possible. When it was needed the Pan-STARRS magnitudes were converted to Johnsons R magnitude using the method described in \citet{Smith2002}. Typical observations consisted of a few individual integrations per night, covering a few hours interval. As our aim with these measurements was to look for long-term variations we produced an average 'per-night' photometric point from the individual integrations which have considerably improved the signal-to-noise ratio, too. These 'per-night' photometry data are presented in Table \ref{tab:phot_opt}. Data obtained with a specific telescope/instrument have been divided into measurement blocks which typically contain data of a few consecutive nights and are separated from the other blocks by a longer period (up to a year). To avoid problems with absolute calibration we allowed a different absolute brightness zero point offset for each of these blocks in the subsequent light curve analysis. Due to the slow apparent motion of Eris heliocentric and observer distance, and phase angle corrections were applied only in those cases when the measurements covered a longer period -- this was the case for some of the literature data \citep[e.g.][]{Rabinowitz2007}, but not for our own measurements which have a typical measurement block length of a few days. In the case of the GROND and the \citet{Rabinowitz2007} data observations were performed by alternating between the $g'$--$r'$ and the $V$--$B$ filters. For our light curve period search we converted the $g'$ data to the $r'$ band using a mean $g'-r'$ colour, and the $B$ data to $V$ data using a mean $B-V$ colour to increase the number of data points for these measurement sequences. Assuming that the spin pole of Eris is coincident with the orbital pole of Dysnomia, and using the pole solution by \citep{Holler2021} we estimated that the aspect angle of Eris' pole changes between $\vartheta$\,$\approx$\,129-133\,deg, i.e. $\Delta \vartheta$\,$\approx$\,4\,deg between the first and last date of the ground-based measurement sequence  (see Table~\ref{tab:other_obs}). 
As the aspect angle is not at its extremes, this does not affect the light curve amplitudes and the detectability of the light curves through different data sets notably.

\subsection{Gaia data \label{sect:Gaia_data}}
\label{sect:gaia}

The Eris Gaia data is available in the third Gaia Data Release \citep{2022arXiv220800211G}, accessible in the Gaia Science Archive\footnote{\url{https://gea.esac.esa.int/archive/}} through the \mbox{{\it gaiadr3.sso\_observation}} table. The table contains
data obtained during the transit of the source on a single CCD, during a single transit. More details about the SSOs in the Gaia DR3 are discussed in 
\citet{2022arXiv220605561T}. 
\begin{table}[H]
    \begin{center}
    \caption{Gaia photometry data of $(136199)$ Eris }
    \label{table:gaiadata} 
    \begin{tabular}{ccc}
        \hline
        Julian Date & $m$ & $\delta m$   \\ 
                    & (mag)    & (mag) \\
        \hline
          2456900.12065   &    -1.21999   &     0.01012 \\
          2456900.12071   &    -1.21999   &     0.01012 \\
          2456900.12076   &    -1.23254   &     0.01313 \\
        ... & ... & ... \\
        \hline
        
    \end{tabular}
    \end{center}
    \footnotesize
{\bf Note.} The table lists the Julian date, brightness (m) and its uncertainty ($\delta$m) in the TESS photometric band. Brightness has been corrected for heliocentric and observer distance, and phase angle.
\end{table}
Gaia G-band data of Eris was corrected for heliocentric and observer distance and phase angle, using spacecraft-centric data obtained from the NASA Horizons system \citep{1996DPS....28.2504G}. We applied a linear phase angle correction using the heliocentric and observer distance corrected brightness values.
We used these reduced magnitudes, as provided in Table~\ref{table:gaiadata} for period search.

\subsection{Period finding method \label{sect:periodfittingmethod}}

We used a residual minimalization method to find the best-fitting light curve period and amplitude in our long-term ground based photometry data (Sect.~\ref{sect:groundbased}).
We chose an amplitude $\Delta m$ and period P, and determined the best fitting light curve phase using a Levenberg–Marquardt minimization algorithm. With these best-fit phase models we calculated the following C(P,$\Delta m$) value for each P-$\Delta m$ pair:
\begin{equation}
    C(P,\Delta m) = \sum_i {\frac{w_i}{N_i}} \sum_j \Bigg( \frac{m^{mod}_{ij}-m^{obs}_{ij}}{\delta m_{ij}} \Bigg)^2
\end{equation}
where the index $j$ denotes the individual, night-averaged values, and $i$ denotes the measurement blocks; $m^{obs}$ and $m^{mod}$ are the measured and model photometry values, respectively, $\delta m$ is the photometric uncertainty, and $N_i$ is the number of individual data points in the measurement block. $w_i$ are the weights of the individual measurements blocks which have been chosen to be $\sqrt{N_i}$. 
We expect that the best-fitting period-amplitude values provide the lowest C(P,$\Delta m$) values. We searched the period range P\,$\in$\,[1d, 17\,d], where the upper limit is set to cover the 15.8\,d is the orbital period of Dysnomia (and it would correspond to a synchronised rotation). The 1\,d lower limit has to be set due to the 'per night' photometry points used in the case of most ground-based measurements. 

\begin{figure}
    \centering
   \includegraphics[width=0.5\textwidth]{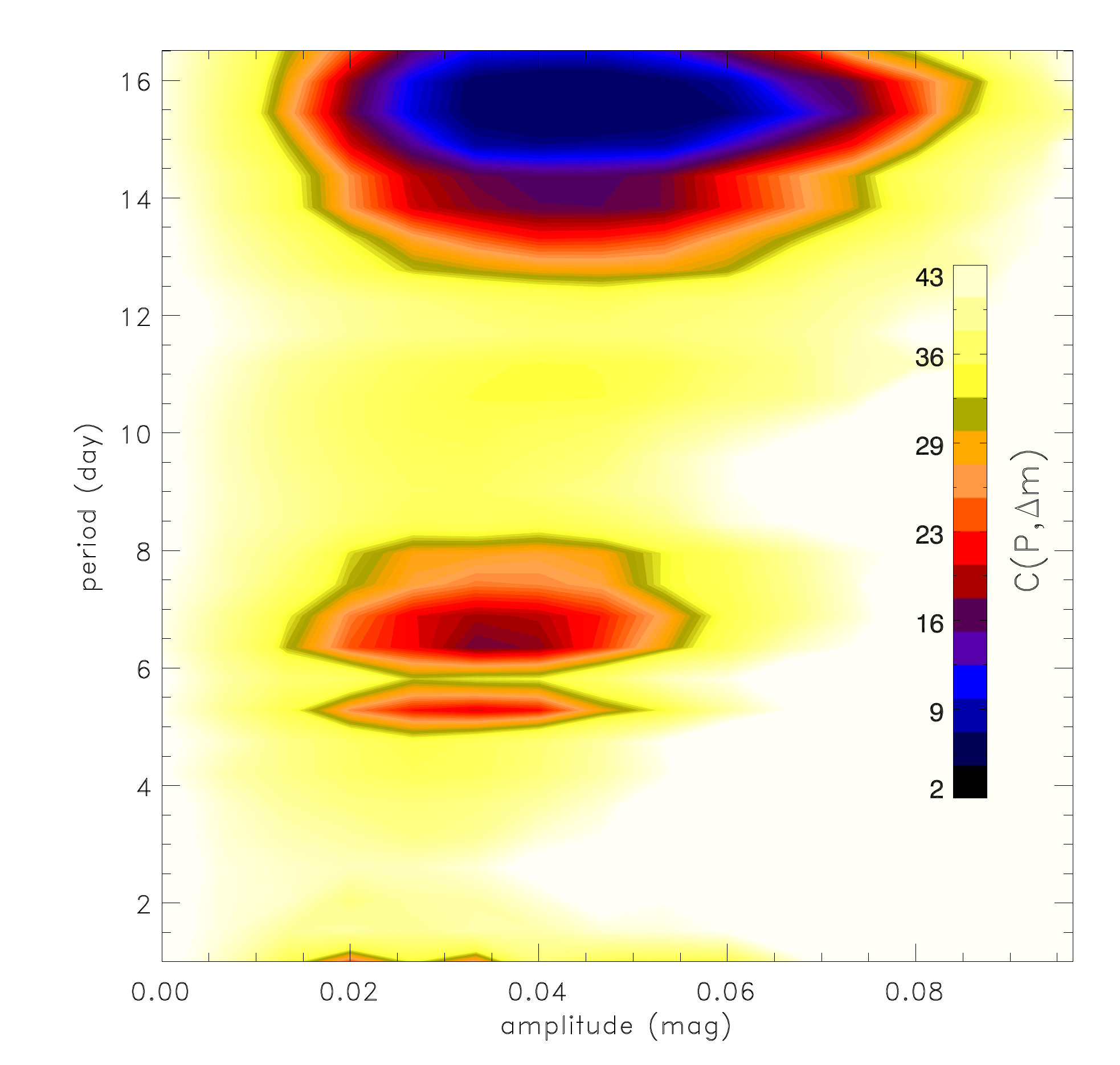}
    \caption{C(P,$\Delta m$) map obtained with the period-amplitude finding algorithm using a synthetic signal with P\,=\,15.78\,d (the orbital period of Dysnomia), as described in detail in the main text.}
    \label{fig:chi2test}
\end{figure}

\label{sect:period}
\normalsize
To check the efficiency of our period/amplitude finding method, we generated a synthetic sinusoidal light curve with a peak-to-peak amplitude of A\,=\,0.040\,mag. We sampled this light curve exactly at the same dates as our real data, and divided these photometry points into the same blocks as the original ones, as described in Sect.~\ref{sect:groundbased}. 
We used the mean photometry error in each measurement block, as assigned a photometric uncertainty to each photometry point in this specific block by assuming a random value with a normal distribution with a standard deviation equal to the mean error. We generated a large sample of synthetic light curves using random light curve phases and 
different random error assignments, and run our period/amplitude finding residual minimalization method. The results show that the period/amplitude can be well recovered with our method, and the expected uncertainty is $\delta$P\,$\approx$\,0.5\,d in the period and $\delta$m\,$\approx$\,0.005\,mag in the light curve amplitude. 
An example of these $\chi^2$ results is shown in Fig.~\ref{fig:chi2test}. 
While the true period and amplitude of the original signal is correctly identified within the uncertainties (dark blue region in Fig.~\ref{fig:chi2test}) there are other, shallower minima popping up at various frequencies, caused by aliasing. E.g. one of these periods is at P\,=\,1\,d due to the single, combined photometry points per night in the case of a number of measurement blocks. 

\begin{figure}
    \centering
   \includegraphics[width=0.45\textwidth]{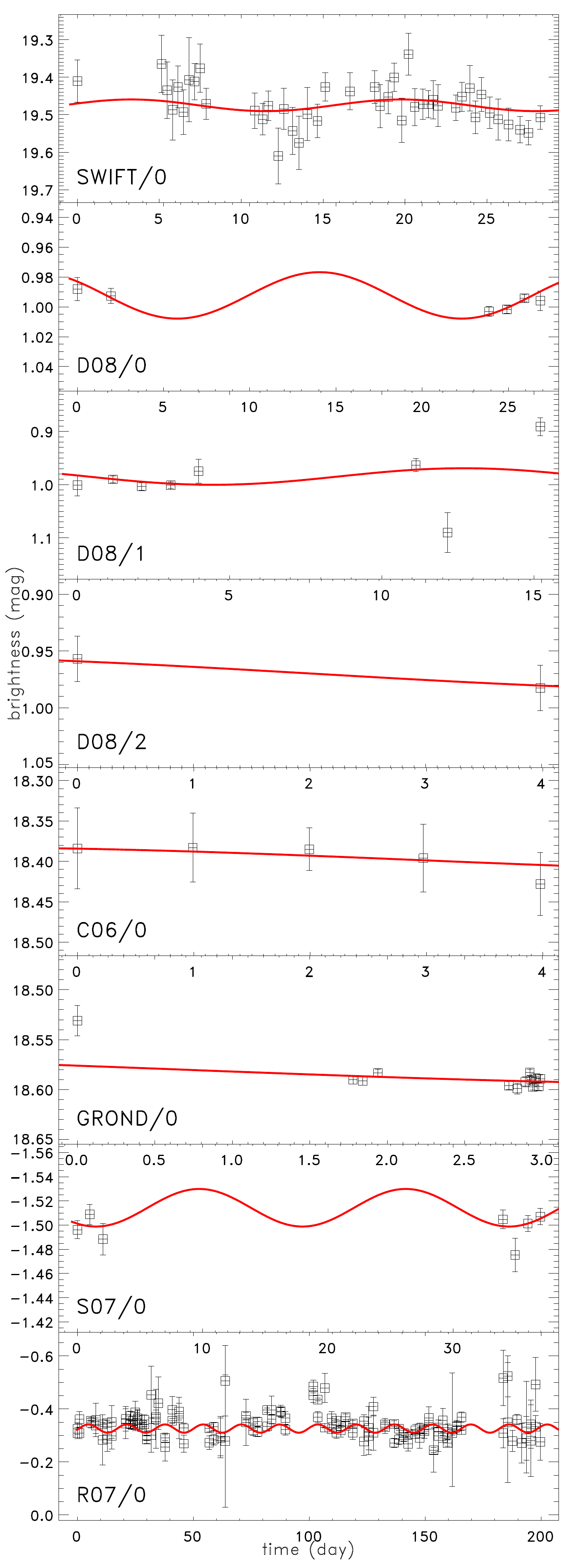}
    \caption{Best matching sine functions (red curve, 15.78\,d, the orbital period of Dysnomia) fitted to the long term photometry data of Eris for some of telescopes considered in our analysis. Time is defined with respect to the start of the measurement block, and brightness is in the actual apparent or reduced system, depending on the telescope and measurement block, as marked in the bottom left corners and described in the text. }
    \label{fig:fitted_sines}
\end{figure}

The match of measurements and the best-fit light curve obtained by the residual minimalization method is shown in Fig.~\ref{fig:fitted_sines} for several measurement blocks.  


\begin{table*}
    \caption{Colours derived from GROND photometry.}
    \begin{center}
    \scriptsize
    \begin{tabular}{ccccccc}
    \hline
    SDSS    & g-r & g-i & r-i & g-z & J-H \\ 
    photometry          & (mag) & (mag) & (mag) & (mag) & (mag) \\ \hline 
    this work & 0.499$\pm$0.004 & 0.623$\pm$0.006 & 0.124$\pm$0.004 & 0.594 $\pm$ 0.013 & -0.479$\pm$0.089  \\ \hline \hline
    Johnson/Cousins  & B-V & V-R & R-I & R-J & J-H   \\ 
    photometry  & (mag) & (mag) & (mag) & (mag) & (mag) \\ \hline 
    this work & 0.709$\pm$0.040 & 0.358$\pm$0.030 & 0.359$\pm$0.02 & 0.64$\pm$0.04 & 0.024 $\pm$0.028 \\ 
     \citet{Carraro2006}& 0.823$\pm$0.023 & 0.391$\pm$0.023 & 0.386$\pm$0.012 &-&-\\
     \citet{Fulchignoni2008}& 0.71$\pm$0.02  &0.45$\pm$0.02 & 0.33$\pm$0.03***&-&-0.29$\pm$0.045*\\
     \citet{DeMeo2009}& -&-&-&-& 0.054$\pm$0.070**** \\ 
     \citet{Perna2010}& -&-&-&-& 0.022$\pm$.070****\\
     \citet{Snodgrass2010} & 0.78$\pm$0.01  & 0.45$\pm$0.03 & 0.33$\pm$0.02  & 0.52$\pm$0.02 &0.287$\pm$0.114**   \\
     \citet{Tegler2016}& 0.75$\pm$0.02& 0.43$\pm$0.02& -&-&-\\
     \citet{AlvarezCandal2020}& 0.782$\pm$0.003  & 0.393$\pm$0.003 & -&-&-\\
     \citet{FernandezValenzuela2021} & 0.74$\pm$0.06  &0.39$\pm$0.05 & 0.38$\pm$0.08***&-&-\\
     \citet{Verbiscer2022} & 0.805$\pm$0.015 & 0.389$\pm$0.049&- &-&-\\ \hline
    \end{tabular}
    \end{center}
    \label{tab:grondcolours}
    \footnotesize
    {\bf Note.} The first row shows the colours directly derived from GROND photometry using the g, r, i, z, J and H bands in this work. Note that the J, H magnitudes are int the AB system. The second raw lists the colours transformed to the B, V, R, I, J, H bands, as described in \citet{Smith2002} and \citet{Blanton2007}. 
    The additional rows represent the values obtained earlier in other studies. *: From \citep{Brown2005} **: \citet{Snodgrass2010} converted to J-H from $J-H_s$ ***: Calculated from V-R and V-I., ****: converted from UKIRT JH to 2MASS JH, using \citep{Cutri2003}
\end{table*}

\subsection{Tidal evolution model}
\label{sect:tidalev}

The satellites of the largest Kuiper belt objects are thought to be formed by large collisions \citep{Barr2016} and remarkably all large Kuiper belt objects with diameters D\,$\gtrsim$\,1000\,km have confirmed satellites \citep{Kiss2017}. The tidal evolution of these systems depends on the size, mass, formation distance, and material properties of the bodies. Tidal evolution has certainly led into a double-synchronous state in the case of the Pluto-Charon system \citep{D97}, but in the Haumea system even the larger satellite Hi'iaka could not reach synchronous rotation \citep{Hastings2016}. Although the spin period of Eris seems to be rather well defined now by this present work, it is an interesting question whether the possible rotation periods of Eris -- ranging from a few hours to the orbit-synchronous state -- are feasible in terms of tidal evolution, using the current knowledge on the system components. This is especially interesting after the likely detection of Dysnomia in the ALMA 870\,$\mu$m data which suggests that Dysnomia could be a massive satellite with a diameter of $\sim$700\,km \citep{BB18}.

We used the simple tidal evolution model by \citep{MD2000}, also used by \citet{Hastings2016}, to calculate the evolution of the satellite orbit (only the semi-major axis in this approximation) and the spin evolution of Eris and Dysnomia. In this model the satellite orbit and the equator of Eris are assumed to be co-planar. Some of the main characteristics of the system originate from the orbit of Dysnomia \citep{Holler2021}, as it defines the current semi-major axis of the satellite orbit, $a_f$, the orbital period, $P_{orb}$, and the system mass, $M_{sys}$. Eris is expected to be nearly spherical and the radius and V-band geometric albedo of Eris, $R_p$\,=\,1163$\pm$6\,km and $p_{Vp}$\,=\,0.96 are known quite precisely from a stellar occultation \citep{Sicardy2011}. The rate of change of the spin frequency depends on the ratio of the second-order tidal Love number and the tidal quality factor, $k_{2p}/Q_p$, the mass and size of Eris, $M_p$ and $R_p$, the actual semi major axis of the satellite orbit $a$, and the mass of Dysnomia, $M_s$:

\begin{equation}
\dot{\omega_p} = -sign(\omega_p - n) \frac{15}{4} \frac{k_{2p}}{Q_p} \frac{M_s^2}{M_p} \bigg( \frac{R_p}{a} \bigg)^3 \frac{G}{a^3}
\label{eq:omegadot}
\end{equation}

The $Q_p$ tidal quality factor was chosen in the range of 10--1000 \citep[][]{Goldreich1966,MD2000} allowing an order-of-magnitude variation around the canonical Q\,=\,100. The second order tidal Love number k$_2$ is calculated from the rigidity $\mu$ following \citet{Hastings2016}. The $\mu_p$ rigidity of Eris was also allowed to vary in a wide range from 10$^9$ to 10$^{11}$\,N\,m$^{-2}$ which should be sufficiently wide to cover typical values from icy to rocky interiors \citep{MD2000}.

The evolution of the semimajor axis can be expressed by the equation below, following \citep{Hastings2016}:
\begin{equation}
a(t) = ({a_f} - {a_0}) (t/T)^{2/13} + {a_0}
\label{eq:at}
\end{equation}

where T is the age of the Solar System, ${a_0}$ is the initial semimajor axis and ${a_f}$ is the present semimajor axis. 

The brightness ratio of Eris to Dysnomia in the F606W band of the HST is 0.0021
\citep{BS07} which defines the size of Dysnomia, $R_s$, for a specific Dysnomia geometric albedo chosen. Among trans-Neptunian objects and satellites a very wide range of albedos are possible from extremely dark surfaces to very bright ones. 
While Dysnomia is probably large and dark \citep{BB18} we choose geometric albedo in the range $p_{Vs}$\,=\,0.02--0.8 for Dysnomia for our model calculations. 
Trans-Neptunian objects with diameters below $\sim$500\,km are expected to have high porosity and low bulk density \citep{Grundy2019}. As a reasonable range we first assumed $\rho_s$\,=\,0.5--2.4\,\gcc, which, together with $R_s$ obtained above, defines the mass of Dysnomia, $M_s$. 
The lower limit is the typical density of objects in the few hundred km size in the Kuiper belt \citep[see e.g.][]{Grundy2019}. 
The upper limit in density, 2.4\gcc\, is the bulk density of Eris, and in extreme cases Dysnomia might have a similar density \citep[as argued e.g. in][]{Holler2021}.   
The mass of Eris is obtained as $M_p$\,=\,$M_{sys}-M_s$, and we use the mass ratio $q$\,=\,$M_s/M_p$ to characterise the system in this sense. 

\begin{figure}
\begin{center}
    \hbox{
    \includegraphics[width=0.42\textwidth]{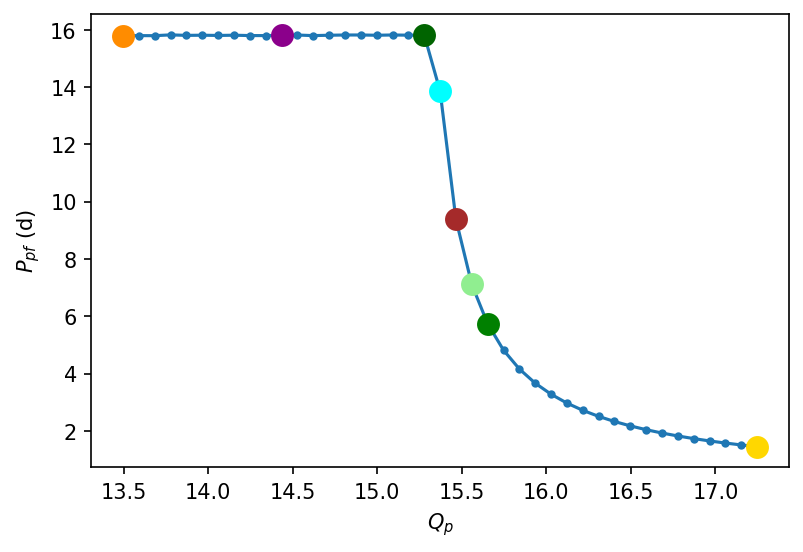}
    }
    \hbox{
    \includegraphics[width=0.42\textwidth]{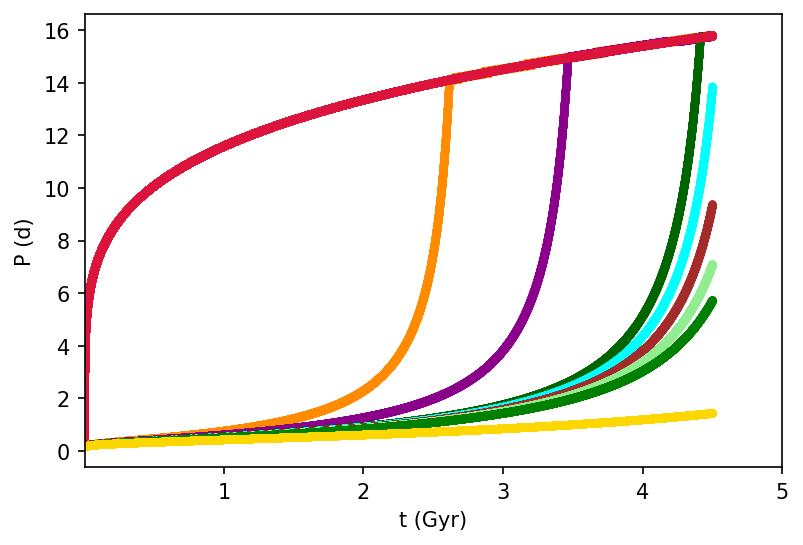}
    }
    \caption{Top panel: Demonstration of the sensitivity of the model to the $Q_p$ parameter. Each dot in this plot corresponds to the $Q_p$ and the final Eris rotation period $P_{pf}$ of a specific run. All other model parameters were kept the same. 
    Bottom panel: Evolution of the rotation period of Eris, using the models presented in the upper panel. The colours of the curves correspond to the colour of the symbols on the upper panel, the red curve is the evolution of the orbital period. }
    \label{fig:qpppmax}
\end{center}
\end{figure}

In addition to the system parameters described above, 
following \citep{Canup2005} the initial semi-major axis of the satellite orbit is assumed to be $a_s$\,=\,1.2$\cdot a_R$, and $a_R$\,=\,$2.456 \cdot R_p \cdot (\rho_p/\rho_s)^{1/3}$ is the Roche limit. The initial spin period of Eris is set by the breakup limit -- 
$\omega_c$\,=\,$(GM_p/R_p^3)^{1/2}$\,=\,$(4\pi G \rho_p/3)^{1/2}$ -- and we allowed spin periods [1--2]\,$\cdot\omega_c$. 

We run the tidal evolution model for a large number of cases assuming independent, random values of $k2_p$,  $Q_p$, $\rho_s$ and $p_{Vs}$ chosen in the intervals described above. We also assume that once Eris has reached synchronous rotation in a simulation run it remains in this state and the rotation period just changes with the changing semi-major axis and orbital period.

In Fig.~\ref{fig:qpppmax} we demonstrate the sensitivity of the model to the $Q_p$ tidal parameter. The model runs presented in this figure led to very different final Eris rotation periods from $P_{pf}$\,$\approx$\,1\,d to synchronized rotation when all starting parameters were kept the same except $Q_p$ which was varied in the range 13.5\,$\leq$\,$Q_p$\,$\leq$\,17.5. 
A very similar sensitivity is seen for the $k_{2p}$ parameter (see Eq.~\ref{eq:omegadot}), however, this parameters is calculated from the $\mu$ or $\mu_{eff}$ values. 

\begin{figure*}[ht!]
\begin{center}
    \hbox{
    \includegraphics[width=0.33\textwidth]{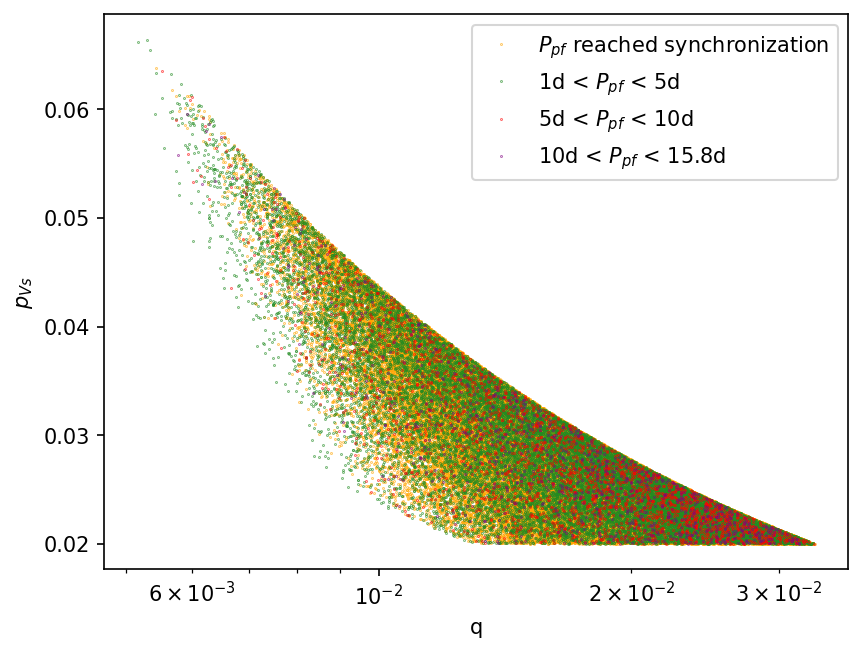}
    \includegraphics[width=0.33\textwidth]{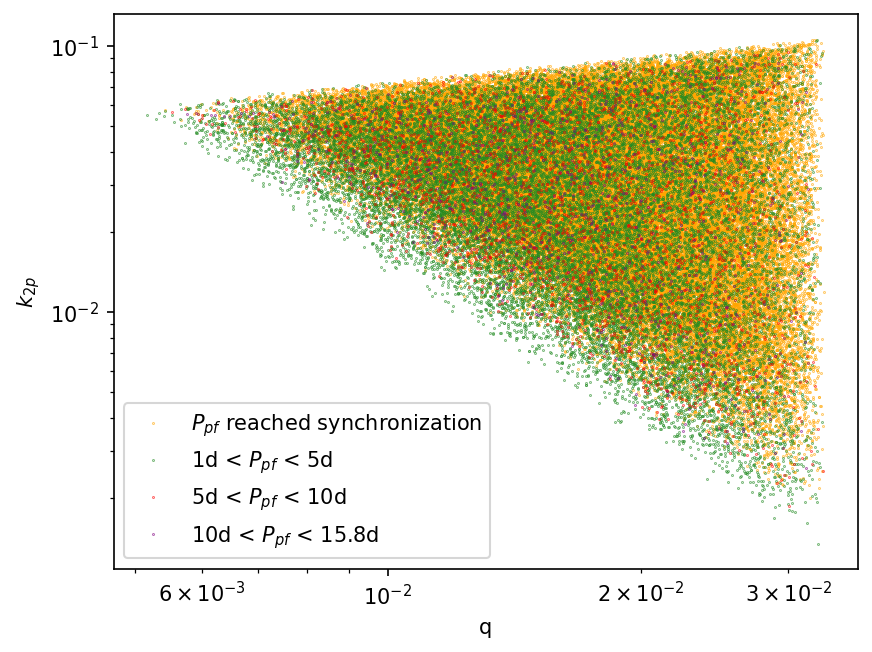}
    \includegraphics[width=0.33\textwidth]{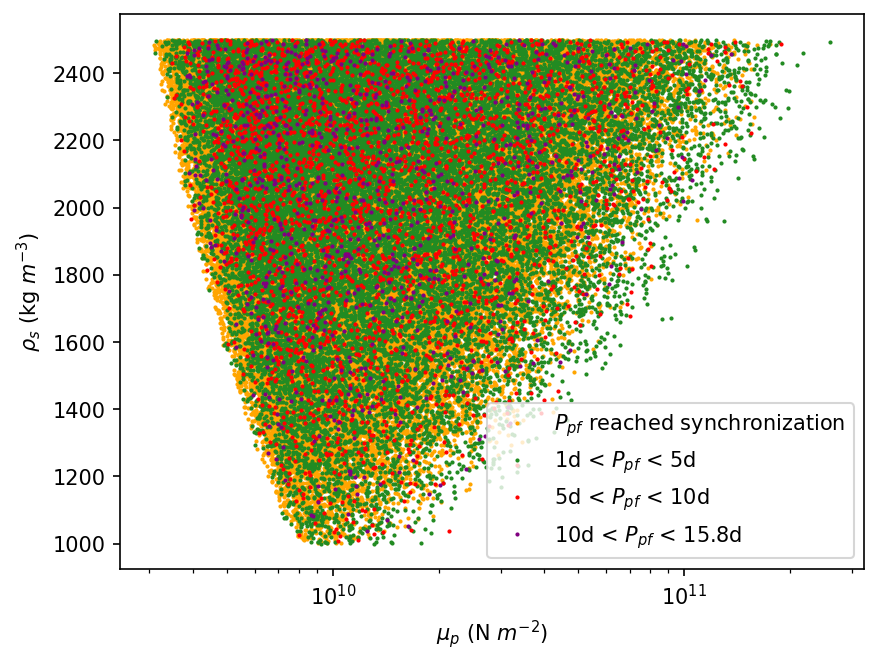}
    }
    \hbox{
    \includegraphics[width=0.33\textwidth]{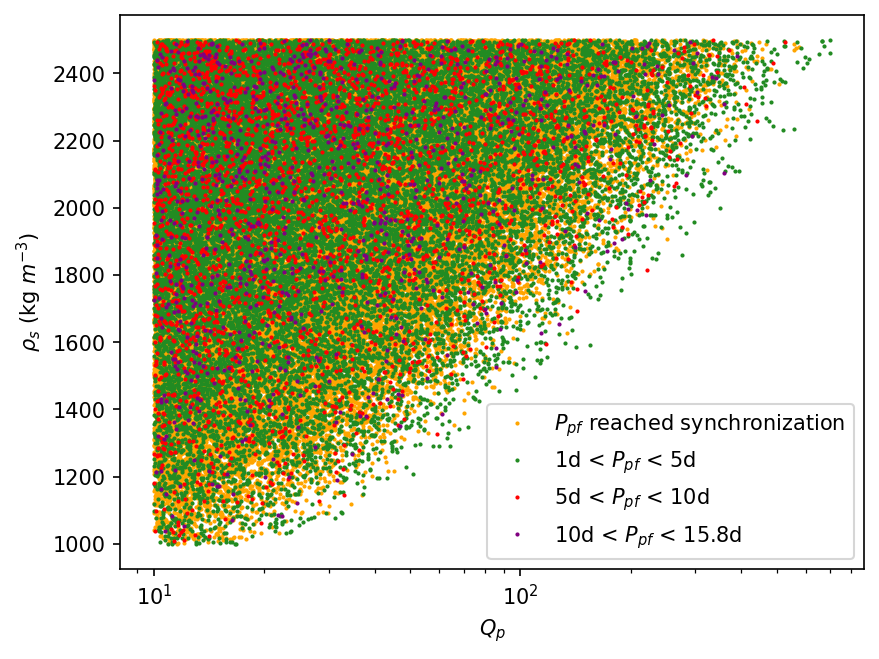}
    \includegraphics[width=0.33\textwidth]{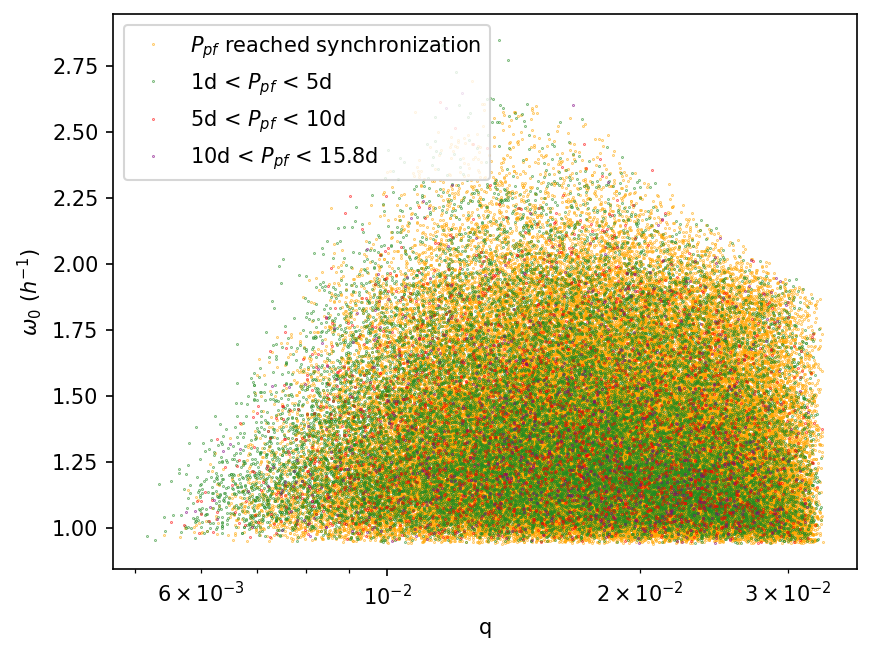}
    \includegraphics[width=0.33\textwidth]{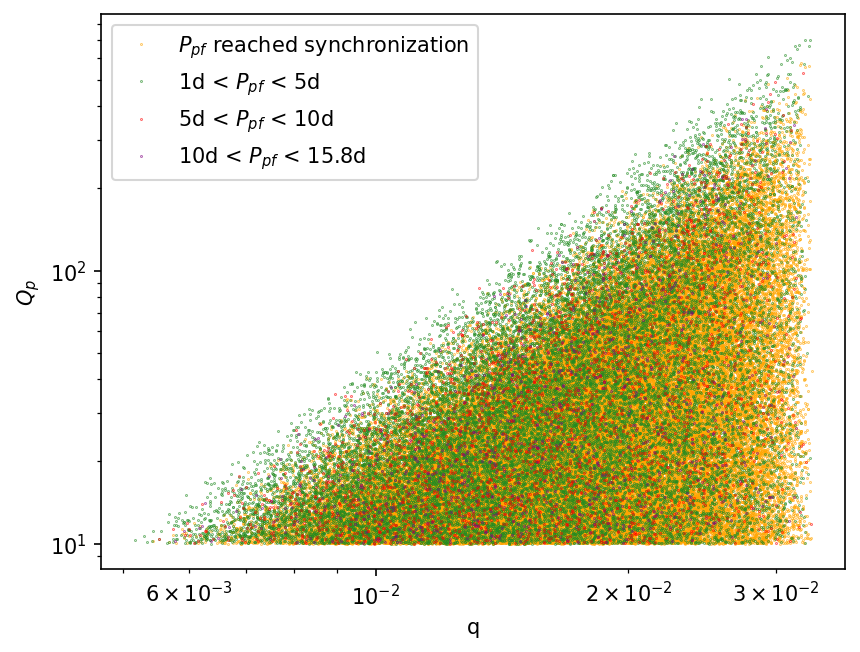}
    }
    \hbox{
    \includegraphics[width=0.33\textwidth]{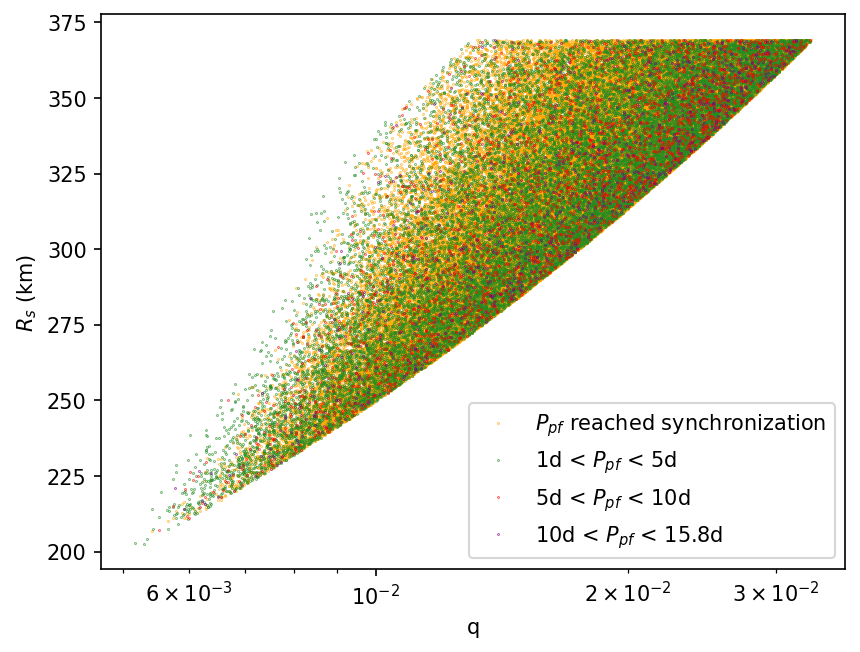}
    \includegraphics[width=0.33\textwidth]{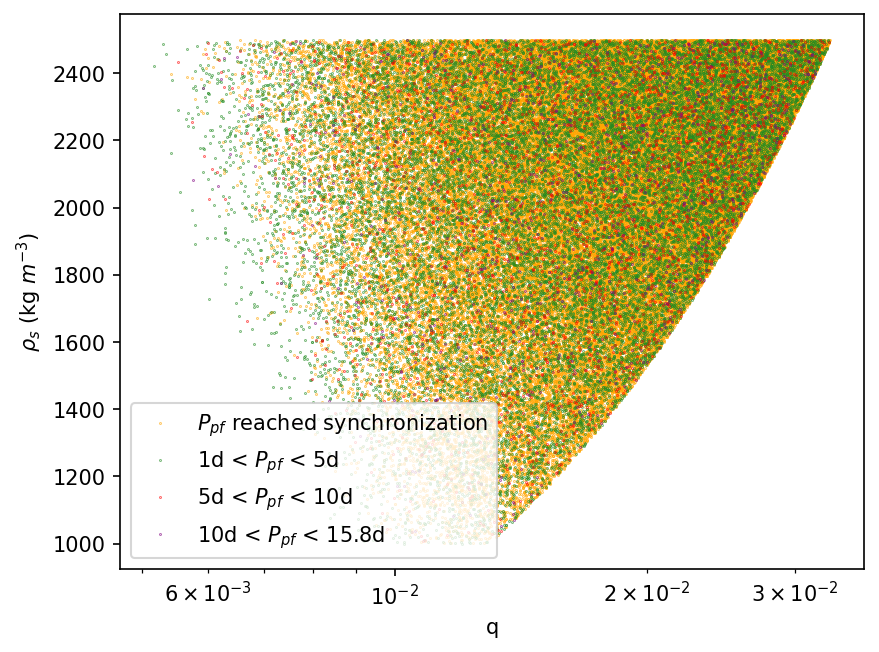}
    \includegraphics[width=0.33\textwidth]{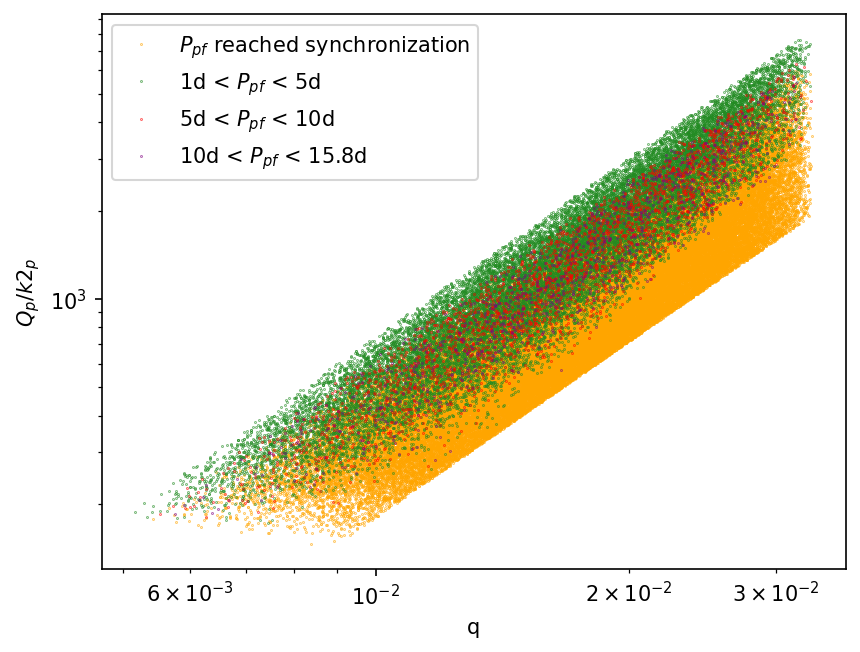}
    }
    \caption{Scatter plots of model parameters for the selected $Q_p$--$M_s$ range cases. The colours correspond to ranges of the final rotation period of Eris ($P_{pf}$), as indicated in the legend boxes.}
    \label{fig:simplots}
\end{center}
\end{figure*}

We also present in Fig.~\ref{fig:simplots} scatter plots of the key parameters of the tidal evolution model (mass ratio $q$; $k_{2p}$ and $Q_p$ tidal parameters, $\mu_p$ rigidity and $\omega_0$ starting angular speed of Eris; density $\rho_s$, albedo $p_{Vs}$ and radius $R_s$ of Dysnomia) which resulted in P$_{pf}$\,$>$\,1\,d final Eris rotation periods. The different colours correspond to ranges of different P$_{pf}$ values or a final synchronized state.

\end{appendix}

\end{document}